\documentclass[twocolumn]{aastex62}
\shorttitle{Extremely Irradiated Hot Jupiters}
\shortauthors{Lothringer et al.}

\usepackage{amsmath}
\usepackage{xspace}
\usepackage{float}

\begin{document}
\title{Extremely Irradiated Hot Jupiters: Non-Oxide Inversions, H$^-$ Opacity, and Thermal Dissociation of Molecules}
\author[0000-0003-3667-8633]{Joshua D. Lothringer}
\affiliation{Lunar and Planetary Laboratory, University of Arizona, Tucson, AZ, USA}

\author[0000-0002-7129-3002]{Travis Barman}
\affiliation{Lunar and Planetary Laboratory, University of Arizona, Tucson, AZ, USA}

\author[0000-0003-3071-8358]{Tommi Koskinen}
\affiliation{Lunar and Planetary Laboratory, University of Arizona, Tucson, AZ, USA}

\vspace{0.5\baselineskip}
\date{\today}
\email{jlothrin@lpl.arizona.edu}





\submitjournal{ApJ}

\begin{abstract}

Extremely irradiated hot Jupiters, exoplanets reaching dayside temperatures ${>}$2000 K, stretch our understanding of planetary atmospheres and the models we use to interpret observations. While these objects are planets in every other sense, their atmospheres reach temperatures at low pressures comparable only to stellar atmospheres. In order to understand our \textit{a priori} theoretical expectations for the nature of these objects, we self-consistently model a number of extreme hot Jupiter scenarios with the PHOENIX model atmosphere code. PHOENIX is well-tested on objects from cool brown dwarfs to expanding supernovae shells and its expansive opacity database from the UV to far-IR make PHOENIX well-suited for understanding extremely irradiated hot Jupiters. We find several fundamental differences between hot Jupiters at temperatures ${>}$2500 K and their cooler counterparts. First, absorption by atomic metals like Fe and Mg, molecules including SiO and metal hydrides, and continuous opacity sources like H$^-$ all combined with the short-wavelength output of early-type host stars result in strong thermal inversions, without the need for TiO or VO. Second, many molecular species, including H$_2$O, TiO, and VO are thermally dissociated at pressures probed by eclipse observations, biasing retrieval algorithms that assume uniform vertical abundances. We discuss other interesting properties of these objects, as well as future prospects and predictions for observing and characterizing this unique class of astrophysical object, including the first self-consistent model of the hottest known jovian planet, KELT-9b.

\end{abstract}


\section{Introduction} \label{section:intro}
There are currently a few dozen known irradiated sub-stellar objects with equilibrium temperatures in excess of 2000 K. Most of these planets are found around A, F, and G type stars with orbital separations of less than 0.05 AU ($a/R_* \lesssim 5$). While some of these planets are the size of terrestrial planets and may or may not have atmospheres \citep[e.g.,][]{lopez:2012,perezbecker:2013b,owen:2013,demory:2016,zahnle:2017}, many of the intensely irradiated objects are extreme versions of hot Jupiters. These planets' short periods, inflated radii, and high temperatures make them some of the most amenable targets to characterize though transit or eclipse spectroscopy and phase curve observations. In fact, using the figure of merit defined by \cite{zellem:2017}, nearly all of the highest signal-to-noise exoplanet targets are ultra-hot Jupiters.

In many ways these planets challenge our understanding of planet formation and evolution. The long standing problem of whether hot Jupiters can form in-situ, and if not, how they migrated to their current location, is made especially acute by short-period hot Jupiters. Extremely irradiated hot Jupiters also have some of the most inflated radii, which interior and evolution models struggle to reproduce \citep{laughlin:2011,thorngren:2017}. Futhermore, and most importantly for this paper, the extreme temperatures found in these planets stretch the capabilities of models built to understand planetary atmospheres with much cooler temperatures. Issues include a lack of important opacity sources present at high temperatures and the lack of consideration of short wavelength irradiation. 
 
Observations of hot Jupiters are commonly interpreted with retrieval techniques to constrain atmospheric properties like the temperature structure and molecular abundance \citep[e.g.,][]{madhusudhan:2009,line:2013b,lee:2013,stevenson:2014,benneke:2015,waldmann:2015,haynes:2015,cubillos:2016,barstow:2017,lavie:2017,evans:2017}. In retrieval analyses, the temperature structures and molecular abundances of spectroscopically important molecules are fit to the data. Several assumptions are typically made in the forward model of the planetary spectrum in order to reduce the explored parameter space and reduce computation time. These assumptions include Local Thermodynamic Equilibrium (LTE), uniform vertical abundances, and limited sets of opacity sources. Such assumptions need to be tested in a self-consistent fashion to help inform the interpretation of retrieval results.

In this paper, we present self-consistent models of extremely irradiated hot Jupiters to provide new insights into the nature of these objects.  Based on this effort, we identify areas where models need improvement or modification and elucidate the path forward toward characterization of these objects.

The rest of Section \ref{section:intro} describes past modeling and observations of hot Jupiters, as well as modeling of irradiated stars, hot Jupiter upper atmospheres, and atmospheric escape on hot Jupiters. In Section \ref{section:model} we describe how we model extremely irradiated hot Jupiters with the PHOENIX atmosphere code. Section \ref{section:results} describes our findings regarding the temperature structures (\ref{section:invs}), molecular abundances (\ref{section:abundances}), opacities (\ref{section:opac}), and the stellar flux penetration depth and contribution functions (\ref{section:pdcf}). We discuss observational implications in Section \ref{section:discussion}, including a look at past observations of extremely hot Jupiters (\ref{section:pastobs}), and conclude in Section \ref{section:conclude}.

\subsection{Previous Modeling of Hot Jupiters and the Effects of TiO and VO}

Much of the early modeling of exoplanet atmospheres focused on the first hot Jupiters discovered, like 51 Peg b, HD 209458b and HD 189733b, planets about 1000~K cooler than what we will consider here \citep[e.g.,][]{burrows:1997,seager:2000,hubbard:2001,barman:2001}. \cite{sudarsky:2003} split hot Jupiters into 5 different classes, with the hottest labeled "roasters" and classified as anything above 1400~K. Planets exceeding 2000~K were soon found and investigations began into the characteristics of these extremely irradiated hot Jupiters. Shortly thereafter, \cite{hubeny:2003} showed the importance TiO opacity has on the temperature structure of planets exceeding equilibrium temperatures of ${\sim} 2000$ K. TiO and VO can provide enough opacity at short wavelengths to heat the atmosphere at pressures of 10-100 mbar, resulting in observable temperature inversions. \cite{fortney:2008} provided a detailed investigation of when TiO and VO opacity becomes important, as well as a discussion on the energetics at play.

Initial analyses of \textit{Spitzer} data seemed to indicate the existence of stratospheres in planets like HD 209458b \citep{knutson:2008}, but it was later shown that the high 4.5 \micron \xspace flux that indicated a temperature inversion was likely due to instrumental systematics \citep{diamondlowe:2014,zellem:2014}. \cite{spiegel:2009}, \cite{knutson:2010}, and later \cite{parmentier:2013} described several processes that may act to remove TiO and VO from the atmosphere of hot Jupiters, preventing temperature inversion from occuring. \cite{spiegel:2009} showed that high vertical mixing is necessary for TiO and VO to persist in the regions of the atmosphere necessary to form temperature inversion and to prevent gravitational settling in regions where TiO and VO may condense in planets like HD 209568b, however vertical cold trapping likely does not play an important role in planets with T$_{eq}>$1900~K \citep{parmentier:2016}. In addition to these vertical cold traps, \cite{parmentier:2013} showed that the nightside of an exoplanet like HD 209458b can act as an effective horizontal cold trap. 
\cite{knutson:2010} proposed the idea that high UV flux, particularly during periods of high stellar activity, may destroy some of the speecies responsible for temperature inversions.

3D global circulation models (GCMs) of the planets in the \cite{sing:2016} sample suggest inversions can form on the dayside of the hottest planets, presumably by TiO and VO but can disappear at the terminator as the influence of irradiation decreases \citep{kataria:2016,wakeford:2017a}. Thus the combination of transit and emission spectra and/or phase curves can provide powerful constraints on the nature of extremely irradiated hot Jupiters.

\cite{molliere:2015} showed that at high temperatures, temperature inversions can form in planets with high C/O ratios. This is due to the fact that the dominant molecule becomes CO rather than H$_2$O. Since CO does not radiate heat as efficiently, the atmosphere is heated around 10 mbar resulting in an inversion of a few hundred Kelvin. This high C/O ratio explanation has been suggested for WASP-18b, which does not show evidence of water absorption or emission in its dayside spectrum but shows CO in emission \citep{sheppard:2017}.

\subsection{Previous Observations of Extremely Irradiated Hot Jupiters}

While it has been shown observationally that exoplanets below ${\sim} 2000$ K likely do not exhibit thermal inversions at the pressures probed by low-resolution near-infrared secondary eclipses, recent discoveries in WASP-18b, WASP-19b, WASP-33b, WASP-121b, and HAT-P-7b show more robust evidence for thermal inversions and/or the presence of TiO \citep{haynes:2015,wong:2016,sheppard:2017,nugroho:2017,evans:2017,arcangeli:2018}. 

Ground-based observations of WASP-19b showed evidence for TiO absorption in the planet's transit spectrum \citep{sedaghati:2017b}. Similarly, a direct detection of TiO emission by \cite{nugroho:2017} in WASP-33b using high-dispersion spectroscopy demonstrated that TiO can indeed exist in exoplanet atmospheres. WASP-33b also shows evidence of a thermal inversion in HST/WFC3 and Spitzer observations \citep{haynes:2015}. H$_2$O and VO emission is suggested in the dayside spectrum of WASP-121b \citep{evans:2017}. Meanwhile, WASP-18b does not show evidence for either H$_2$O, TiO, or VO emission or absorption in its inverted atmospheres, but the large dayside flux measured at 4.5 microns may be evidence of CO emission, characteristic of a thermal inversion \citep{sheppard:2017,arcangeli:2018}. HAT-P-7b also has large 4.5 micron flux, hinting at a thermal inversion in that planet as well \citep{wong:2016}.

WASP-12b has attracted controversy over whether it exhibits a temperature inversion or not. Spitzer photometry at 3.6 and 4.5 microns points to molecular absorption \citep{stevenson:2014}, implying no temperature inversion, but photometry at other wavelengths suggests an isothermal or weakly inverted atmosphere \citep{cowan:2012,crossfield:2012d}. HST/WFC3 eclipse spectra of WASP-12b show no evidence for H$_2$O emission or absorption, also suggesting an isothermal atmosphere at pressures probed by water \citep{swain:2013}, though this has also been used to argue for a high C/O ratio \citep{stevenson:2014}. H$_2$O has been detected in the transit spectrum of WASP-12b and retrievals that assume chemical equilibrium constrain the C/O ratio to be ${<}$1 \citep{kreidberg:2015}.

A handful of extremely irradiated planets do not appear to have temperature inversions at the pressures sensed by secondary eclipse observations, namely Kepler-13Ab and KELT-1b. Water absoption in the 6.5 M$_{Jupiter}$ Kepler-13Ab points towards a monotonically decreasing temperature structure \citep{beatty:2017a}, while spectrally resolved H-band measurements of the 27 M$_{Jupiter}$ KELT-1b also support a non-inverted scenario \citep{beatty:2017b}. Surface gravity may play a role in preventing an observed inversion in these planets by improving cold trap efficiency \citep{beatty:2017b}.

\subsection{Irradiated Stars and Brown Dwarfs}

Planets are not the only companions to experience intense irradiation. Both brown dwarfs and stars can be close enough to a hot primary body for irradiation to change the secondary's atmosphere significantly. Studies of irradiated M-dwarfs orbiting white dwarfs find that large temperature inversions exist in the secondary's atmosphere and many of the molecules that exist in non-irradiated M-dwarf atmospheres have been thermally dissociated \citep{brett:1993,barman:2004}. 

A handful of brown dwarfs also orbit white dwarfs and experience intense irradiation \citep{burleigh:2006,casewell:2015,santisteban:2016}. Two of these brown dwarfs, WD0137-349B and EPIC212235321B, exhibit emission from metal lines, suggesting a chromosphere-like temperature inversion in its atmosphere \citep{longstaff:2017,casewell:2018}. Our present investigation of extremely irradiated exoplanets is directly applicable to these other classes of irradiated objects.

\subsection{The Upper Atmosphere and Atmospheric Escape on Hot Jupiters}

Atmospheric escape has been observed on a handful of hot Jupiters to date, namely HD209458b, HD189733b, and WASP-12b \citep{vidal-madjar:2003,etangs:2010,etangs:2012,fossati:2010}.  Models of more moderately irradiated hot Jupiters HD209458b and HD189733b indicate that atmospheric escape does not drastically alter the total mass of these planets throughout the planet's lifetime \citep[e.g.,][]{yelle:2004,yelle:2006,murray-clay:2009,koskinen:2012a,chadney:2017}. The same does not necessarily hold for extreme hot Jupiters such as WASP-12b that undergo significant Roche lobe overflow in addition to thermal escape due to their close orbit around the host star \citep{li:2010,jackson:2017}. Also, the temperatures found in the lower and middle atmospheres of the most extremely irradiated hot Jupiters are similar to the temperatures in the upper atmospheres or thermospheres of the more moderately irradiated hot Jupiters. This is likely to further enhance mass loss rates from extreme hot Jupiter atmospheres.

Models and observations of moderate hot Jupiters indicate that their atmospheres undergo hydrodynamic escape roughly at the energy-limited rate, which depends linearly on the heating efficiency of the upper atmosphere \citep{watson:1981,yelle:2004,garcia-munoz:2007,murray-clay:2009}. In contrast to Jupiter where only a small fraction of molecular hydrogen in the thermosphere is dissociated by solar UV radiation, almost all of the H$_2$ dissociates in the upper atmospheres of hot Jupiters \citep{coustenis:1998,yelle:2004,yelle:2006}. Recent models indicate that a combination of thermal dissociation and water dissociation chemistry leads to the dominance of atoms and ions at ${\lesssim}1$ microbar in the upper atmosphere of HD209458b \citep{moses:2011,koskinen:2012a}. The lack of effective radiative cooling above the dissociation front allows the thermosphere to reach a peak temperature of about 10,000~K.  Evidence for this temperature inversion in the upper atmosphere has been obtained from observations of the sodium resonance doublet at 5890 and 5900 \AA \xspace on HD209458b, HD189733b and WASP-49b \citep{vidalmajar:2011,wyttenbach:2015,wyttenbach:2017}. Once hydrodynamic escape sets in, the temperature decreases with altitude above the heating peak due to adiabatic cooling from the expansion of the atmosphere. The resulting escape rate is typically high enough to drag heavier oxygen, carbon, magnesium, and silicon atoms out of the atmosphere and these species are also detectable in transit observations \citep{vidal-madjar:2004,vidal-madjar:2013,linsky:2010,fossati:2010,koskinen:2012a}.
	   
Based on the mechanism outlined above, \cite{koskinen:2007} and \cite{koskinen:2014} argued that thermal hydrodynamic escape occurs only if the stellar X-ray and UV (XUV) flux is sufficient to dissociate molecules. Most lower mass hot Jupiters fall into this category while higher mass planets such as WASP-18b undergo much slower kinetic (Jeans) escape even at very close-in orbits where the upper atmosphere is composed of atoms and ions \citep{fossati:2018}.  Lower mass extreme hot Jupiters present an interesting test case for models of atmospheric escape.  The high temperatures in their atmospheres, not limited to the thermosphere, dissociate molecules deeper than on moderate hot Jupiters, and can lead to rapid escape enhanced by Roche lobe overflow. KELT-9b, the hottest known Jovian exoplanet (T$_{dayside}$ = 4,600 K), is particularly interesting.  \cite{gaudi:2017} estimated a range of mass loss rates for this planet based on the energy-limited formalism. Their upper limit on the mass-loss rate implied that the planet would lose its entire atmosphere in only 600 Myr, similar to the main-sequence lifetime of its A0-type host star. The atmosphere models that we present here will provide useful lower boundary conditions for detailed escape models of extreme hot Jupiter atmospheres that bear on their formation history and long-term evolution.

\section{Methods} \label{section:model}
We model extremely irradiated hot Jupiters using the PHOENIX atmosphere code, Version 16.10 \citep{hauschildt:1997,hauschildt:1999b} with irradiation \citep{barman:2011,barman:2005,barman:2007}. This code solves for radiative-convective equilibrium iteratively with chemical equilibrium, such that flux is conserved at each layer. Models are started with an isothermal temperature profile near the equilibrium temperature of the planet, after which the model iterates on the temperature structure via the Uns\"old-Lucy method \citep{lucy:1964,hauschildt:2003}, calculating chemical equilibrium and radiative transfer at each iteration, until flux conservation is achieved. We also investigated different starting conditions to confirm results.

Radiative transfer is calculated line-by-line in plane parallel geometry using accelerated $\Lambda$-iteration \citep{hauschildt:1999b}. The model is calculated on an optical depth grid of 64 layers evenly spaced in log-space from $\tau = 10^{-10}$ to $10^{2}$ at 1.2 microns. For most of our models, this corresponds to pressures of $10^{-12}$ to ${\sim}$50 bars. Note that at pressures below $10^{-6}$ bar, some NLTE processes that we do not include may become important. Both the planetary and stellar spectrum are calculated from 10 to 10$^6$ \AA \xspace (0.001-100 microns). The models include opacity from 130 molecular species, including many isotopes and deuterated molecules, and atomic species up to uranium, including many ionized states. PHOENIX also takes into account many continuous opacity sources, including bound-free (i.e., photoionization) opacity from H, H$^-$, He, C, N, O, Na, Mg, Al, Si, S, Ca, and Fe, free-free opacity from H, Mg, and Si, and scattering from $e^-$, H, He, and H$_2$. Collision induced absorption (CIA) from H$_2$ collisions with H$_2$, He, Ar, CH$_4$, and N$_2$, as well as CH$_4$-CH$_4$, CO$_2$-CO$_2$, and Ar-CH$_4$ CIA are included.

Chemical equilibrium is calculated using the Astrophysical Chemical Equilibrium Solver (ACES) using 894 different species in the equation of state, including 83 different elements up to atomic number 92, uranium. While photoionization cross-sections are included in the opacity calculation, they are not self-consistently included in the chemical equilibrium solution. Thus all ionization that occurs in our models is due solely to thermal ionization.

\subsection{Fiducial Model}

We explore a number of models for a generic extremely irradiated hot Jupiter. This generic planet serves as a fiducial example to investigate general properties of these planets. We use solar metallicity, a mass of 1 M$_{Jupiter}$, and an inflated radius of 1.5 R$_{Jupiter}$, similar to a lower mass WASP-33b, for comparison. The planet orbits an F0 star with an effective temperature of 7200 K at 0.025 AU. We also vary our generic hot Jupiter's orbital radius between 0.025 AU and 0.1 AU, effectively varying the planet's equilibrium temperature between 1600 and 3200 K. All models assume uniform heat redistribution across the entire dayside (i.e., the outgoing flux radiates over 2$\pi$ steradians), unless otherwise noted. Models with full planet-wide heat redistribution would have temperatures about 400-500 K cooler.

\begin{figure*}[ht]
	\center    \includegraphics[width=7in]{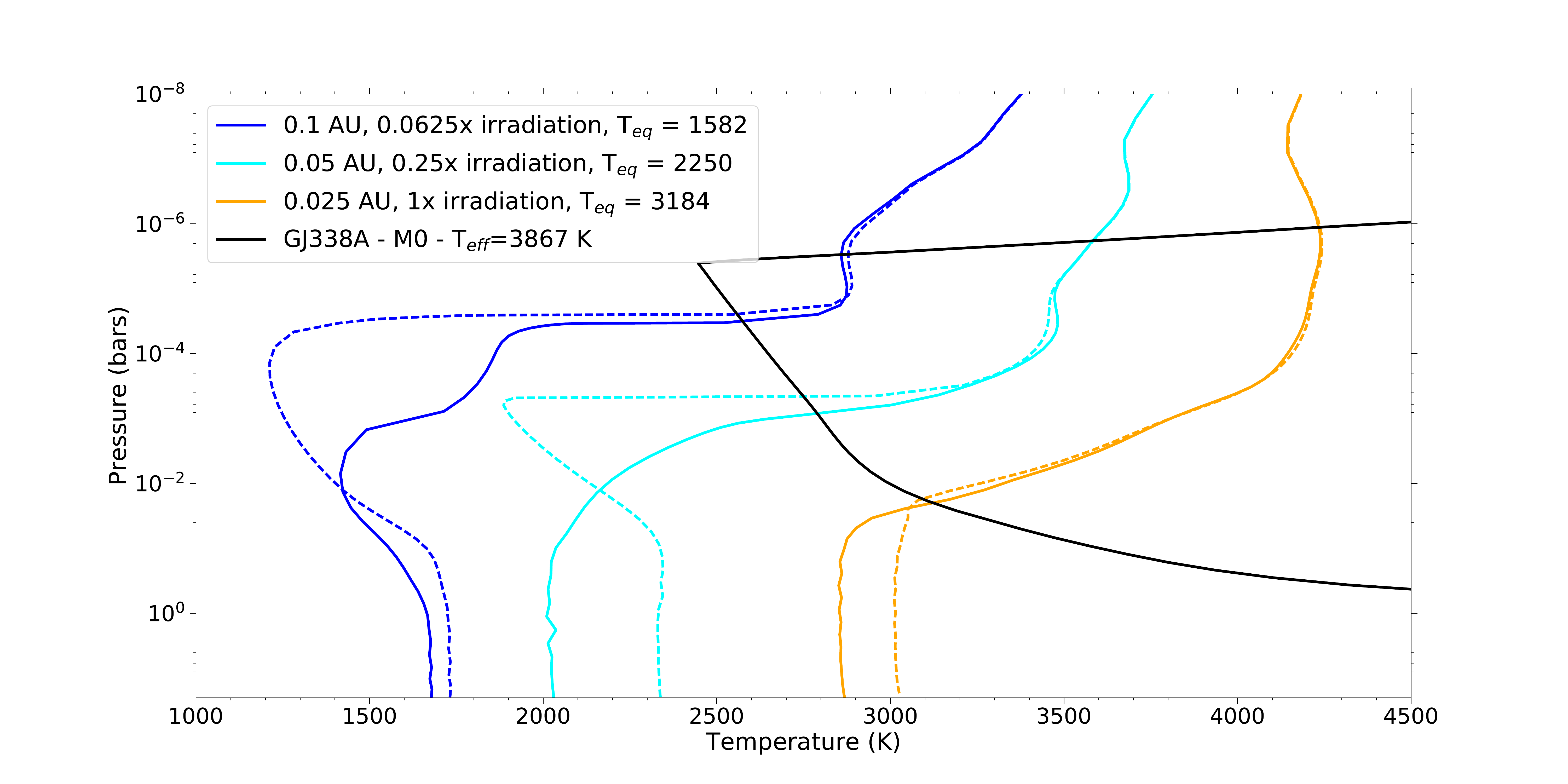}
	\caption{Pressure-temperature profiles of our fiducial hot Jupiter at different orbital separation. Solid lines are models with TiO and VO and dotted lines are models without TiO and VO. The pressure-temperature profile of a M0 dwarf star with a chromosphere is also overplotted from Peacock et al. (in prep). \label{fig:tp_sma}}
\end{figure*}

\section{Results} \label{section:results}

\subsection{Temperature Inversions} \label{section:invs}

\begin{figure}[ht]
	\center    \includegraphics[width=3.5in]{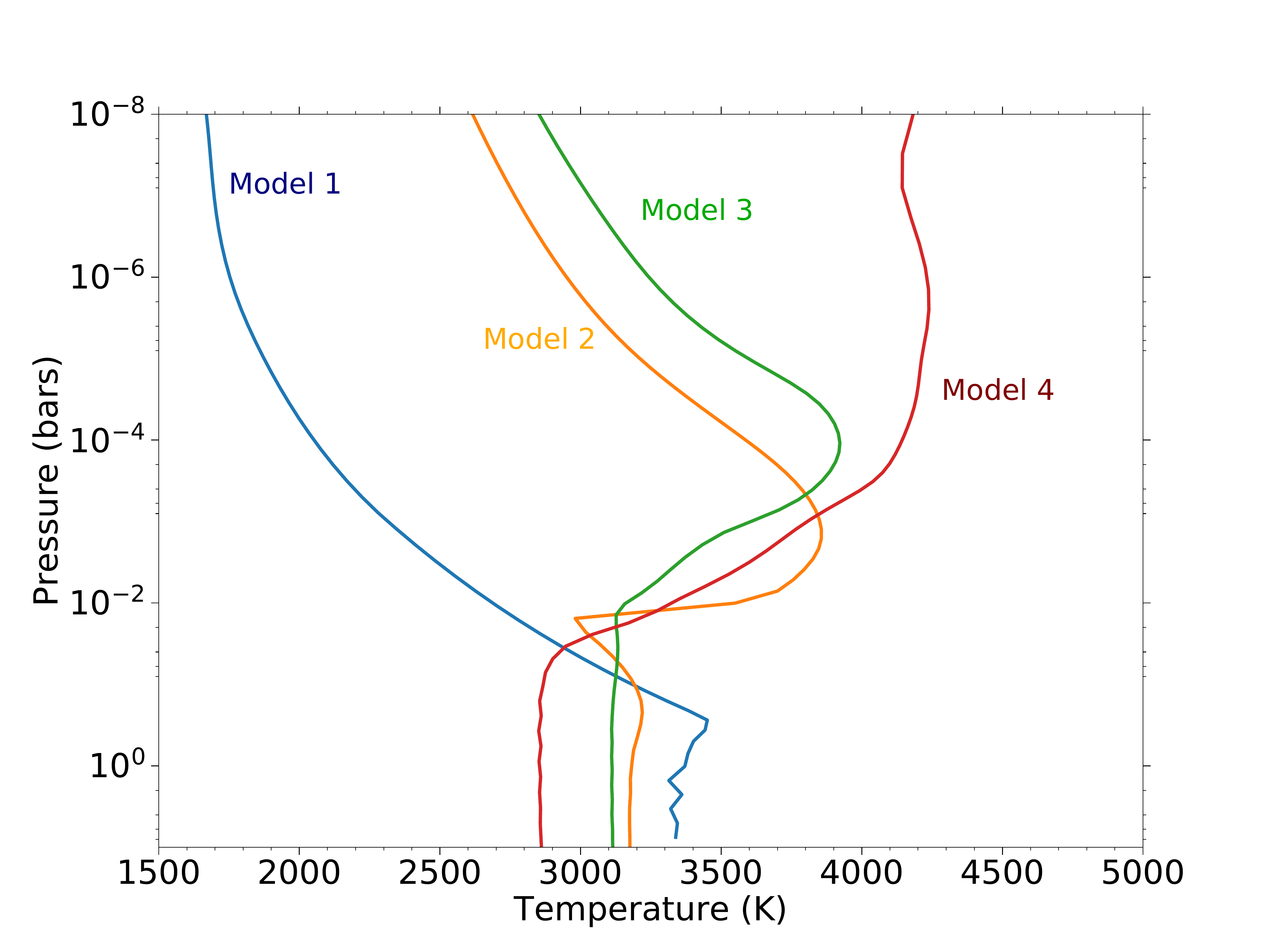}
	\caption{Pressure-temperature profiles of our generic hot Jupiter at 0.025 AU when including different opacity sources. All models in this figure do not have TiO or VO opacity. Model 1 (blue) shows only H$_2$-H$_2$ continuous opacity, atomic absorption from H, He, Na, and K, and molecular absorption from only major hot Jupiter absorbers common in cooler planets like H$_2$O, CO, and CO$_2$. Model 2 (gold) is the same as Model 1 but contains opacity from SiO and metal hydrides. Model 3 (green) is the same as Model 1 but also contains Fe atomic absorption and full continuous opacity (including H$^-$ and bound-free Fe opacity). Model 4 (red) contains all opacity sources available in PHOENIX. \label{fig:fe_tp}}
\end{figure}

Figure \ref{fig:tp_sma} shows pressure-temperature profiles of our generic hot Jupiter at several orbital separations. Relatively far away from its host star, at 0.1 AU, the planet has an equilibrium temperature of about 1600 K and exhibits no inversion near the pressures probed by secondary eclipse or transit observations (${\sim}$1 mbar to 1 bar); however, a thermosphere at pressures below 1 mbar does exist due to the absorption of high-energy UV radiation. Both models with and without TiO and VO have decreasing temperatures with altitude up to about a mbar due to the fact that TiO and VO remain mostly condensed (see Figure \ref{fig:tio_pp} and \ref{fig:vo_pp}). The radiative-convective boundary (RCB) is outside of the region of the atmosphere we model, implying that the the RCB occurs at pressures $\geq$50 bar and where $\tau > 100$. This is consistent with previous theory showing that the RCB is pushed to deeper pressures in irradiated objects \citep{guillot:2002,parmentier:2015}.

As we move the planet closer to its host star to 0.05 AU, we can essentially see the thermosphere move further up in pressure as the irradiation has increased by a factor of 4 and the equilibrium temperature has increased to 2250 K. At 1 mbar, the planet is as hot as an M0 dwarf at the same pressure. Additionally, in the models including TiO and VO, these species have evaporated to the gas phase and become an important opacity source, causing a temperature inversion below 0.1 bars, as in \cite{fortney:2008}. At these temperatures there is a dichotomy of atmospheres with and without TiO and VO on their dayside.

Even closer to the star at 0.025 AU, the planet has an equilibrium temperature of about 3200 K. Temperature inversions are present regardless of whether TiO or VO are included in the model. In some sense, the thermosphere that was at 1 mbars when the planet was at 0.1 AU is now at 0.1 bars, pressures that are probed with secondary eclipse and transit observations. Thermospheres are found in all solar system planetary atmospheres; extremely irradiated hot Jupiters are unique in that their thermospheres occur at pressures important for the thermal emission of the planet (i.e., near the maximum of the planet's near-infrared contribution function, see Section \ref{section:pdcf}).

We find that a number of factors contribute to this strong inversion at 0.025 AU. First, atomic metal opacity is capable of absorbing enough short wavelength irradiation to heat up the atmosphere. Figure \ref{fig:fe_tp} shows that the addition of Fe opacity with full continuous opacity treatment is enough to create a thermal inversion at 10 mbars. The bound-bound opacity of Fe absorbs significantly longward of 0.3 microns, where the irradiation from the host star peaks. Additionally, the bound-free opacity absorbs the high-energy flux shortward of 0.3 microns \citep{sharp:2007}. Other atomic opacity sources, primarily the other metals like Mg and Si, help to increase this effect even more. The addition of other important molecules besides TiO and VO will also create an inversion. These molecules include SiO and the metal hydrides, all of which absorb efficiently at short wavelengths \citep{sharp:2007}. We discuss the opacity structure of the atmosphere more in Section \ref{section:opac}.

Some previous modeling has also pointed out the possibility of non-oxide driven inversions. As mentioned above, \cite{brett:1993} and \cite{barman:2004} showed that dramatic temperature inversions can occur in the atmospheres of M-dwarfs irradiated by white dwarfs with temperatures too hot for TiO to form. Also described above, high C/O atmospheres can have temperature inversions caused by a lack of molecules like H$_2$O to radiatively cool the atmosphere \citep{molliere:2015}.

We found it difficult to create non-inverted atmospheres at these high temperatures. In order to create the non-inverted profile in Figure \ref{fig:fe_tp}, we had to remove a number of opacity sources, with the only remaining opacity sources being atomic opacity from H, He, and the alkali metals, molecular opacity from H$_2$O, CO, CO$_2$, CH$_4$, H$_2$S, H$_2$, HCN, NH$_3$, OH, and PH$_3$, and continuous opacity from H$_2$-H$_2$ CIA. These opacity sources are often assumed to be enough to describe the atmosphere of lower temperature hot Jupiters. We find here that additional opacity sources are necessary to adequately model extremely irradiated hot Jupiters.

\begin{figure*}[p!]
	\center    \includegraphics[width=7in]{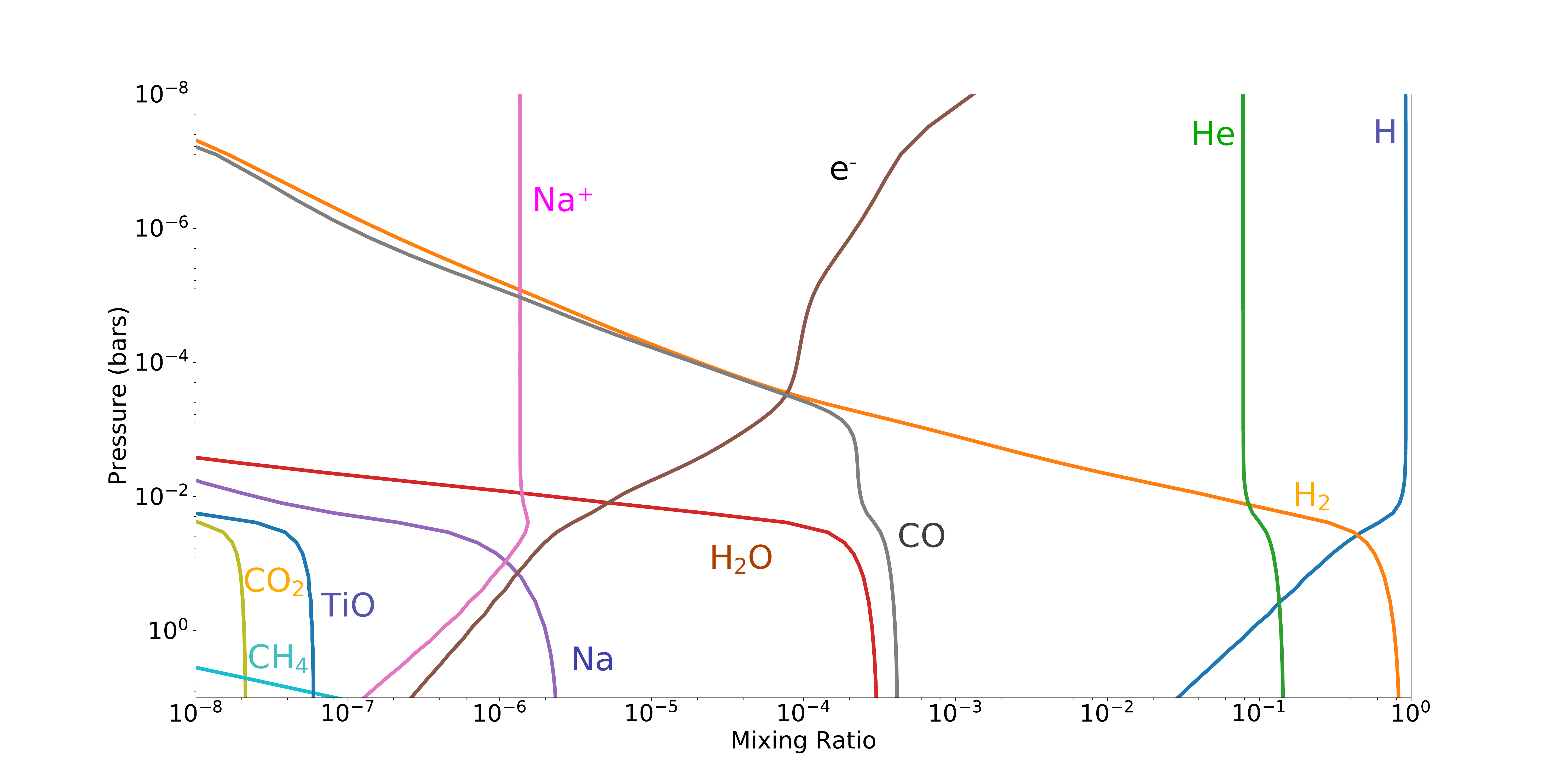}
	\caption{Mixing ratios in chemical equilibrium as functions of pressure for different species commonly studied in hot Jupiters for our generic hot Jupiter at 0.025 AU. Most neutral atoms and molecules are depleted at pressures probed with near-infrared secondary eclipse spectra. Particularly important is the H$_2$ dissociation altitude, which occurs around 10 mbar. Similarly, Na is mostly ionized near 100 mbar. Electron mixing ratios reach $10^{-4}$ at about 1 mbar. Photoionization is not included in the chemical equilibrium solution so all ionization in due to thermal ionization. \label{fig:pp}}
\end{figure*}

\begin{figure*}[ht]
	\center    \includegraphics[width=7in]{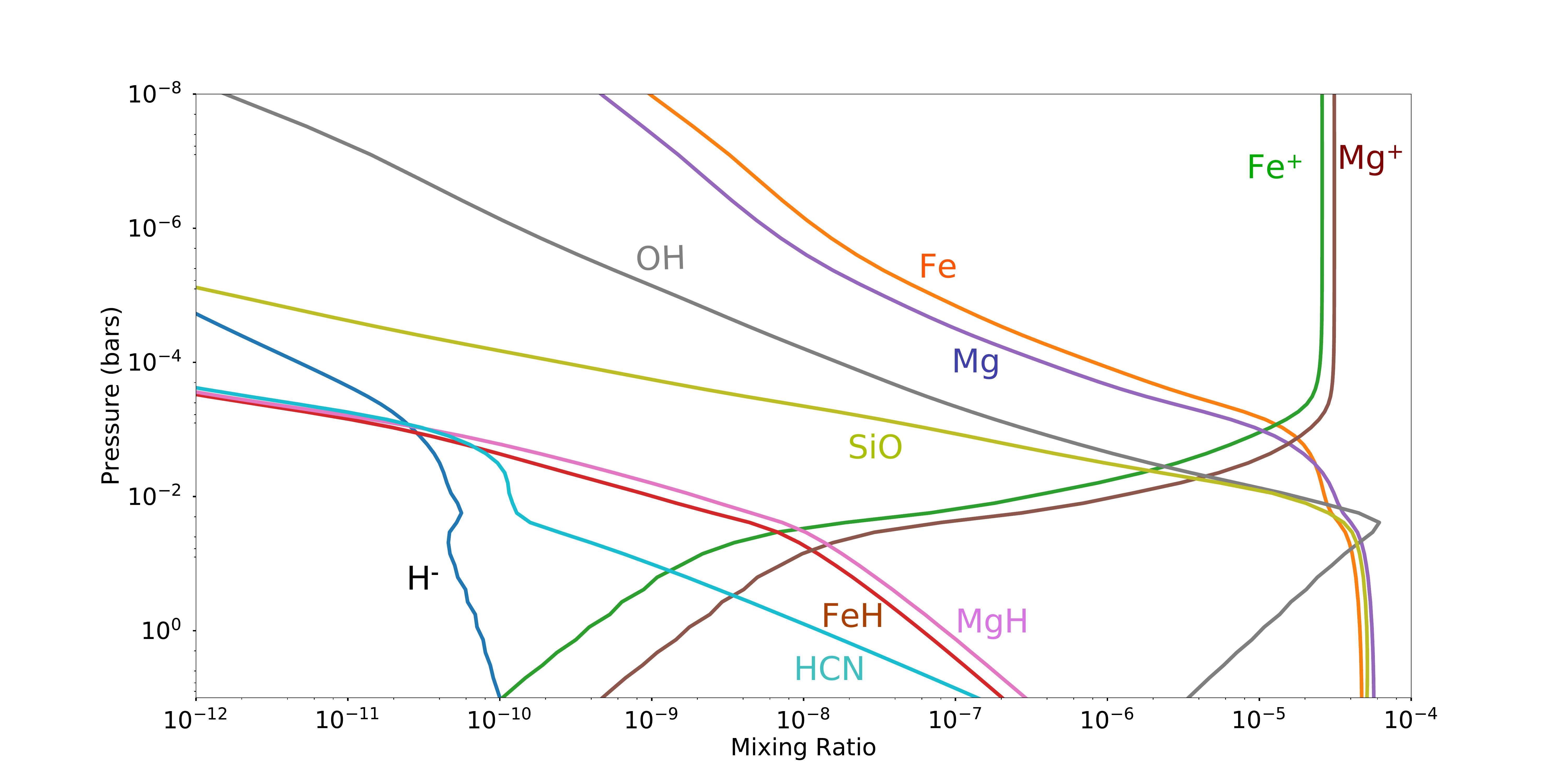}
	\caption{Same as Figure \ref{fig:pp} but for other species that are important opacity sources at high temperatures.
		\label{fig:pp_trace}}
\end{figure*}

\subsection{Atomic and Molecular Abundances} \label{section:abundances}

While the atomic metals are absorbing the short wavelength flux from the host star, molecules that are responsible for radiative cooling  atmosphere do not exist due to the extreme temperatures. Figures \ref{fig:pp} and \ref{fig:pp_trace} show the mixing ratio of important atomic and molecular species in the atmosphere of our generic hot Jupiter at 0.025 AU. H$_2$O becomes heavily depleted due to thermal dissociation below 10 mbar, while CO$_2$ and TiO become depleted by 50 mbar. CH$_4$ is not in abundance below 10 bars. Thus below 10 mbars, the only molecule in abundance is CO, being held together by its triple bond, the strongest in nature. CO, however, is not an efficient coolant because its roto-vibrational spectrum is confined to a single vibrational mode. The combined effects of effective short wavelength absorption and poor long wavelength cooling lead to strong thermal inversions. For full day-to-night temperature redistribution, molecules can survive about an order of magnitude lower in pressure. 

\begin{figure}[t]
	\center    \includegraphics[width=3.5in]{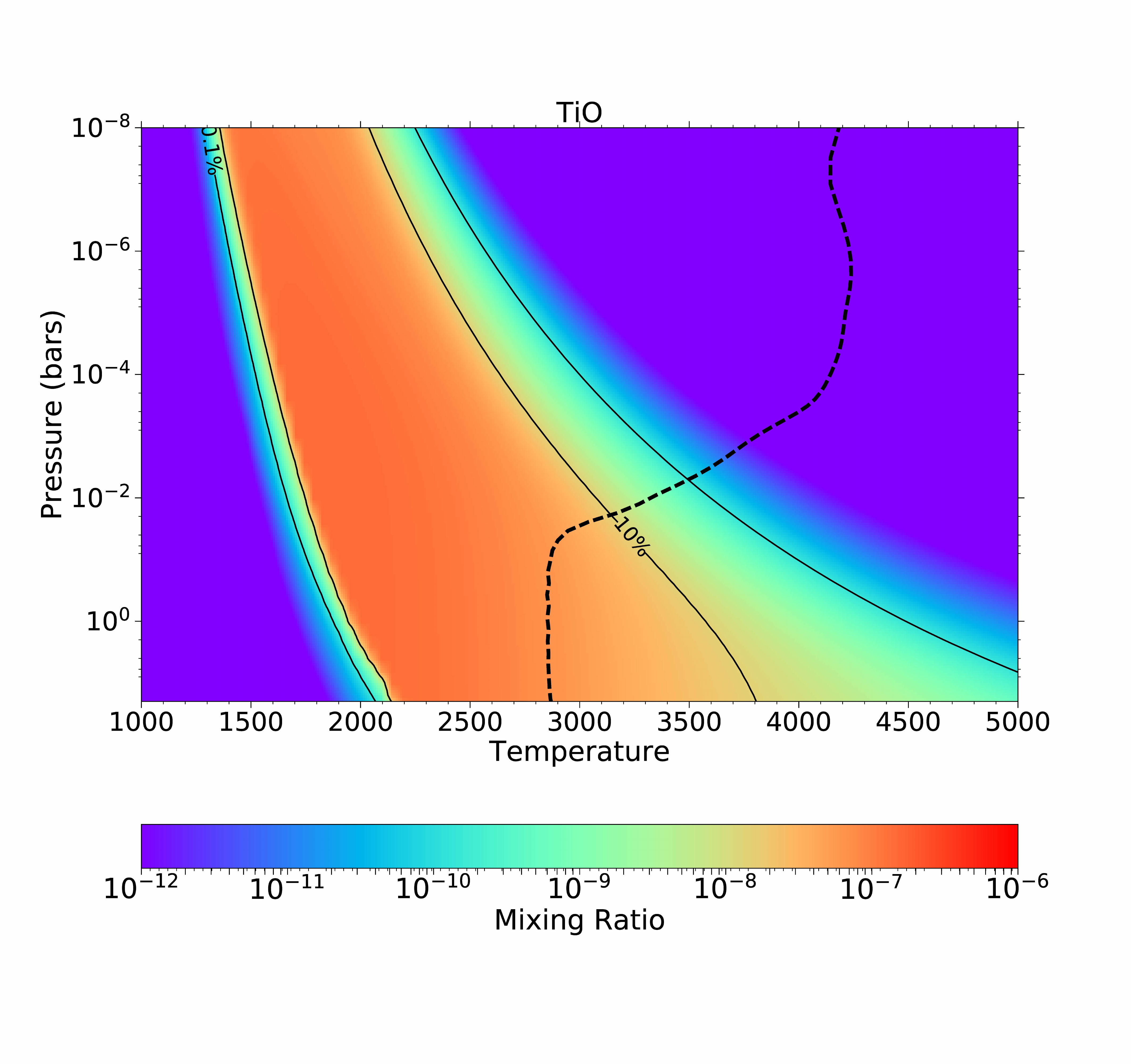}
	\vspace{-40pt}
	\caption{Mixing ratio of gaseous TiO versus temperature and total pressure in chemical equilibrium with all other species considered in PHOENIX. Areas of red color indicate high TiO abundances. At low temperatures, TiO has condensed out of the gas phase, so gaseous TiO abundances are low. At high temperatures, TiO becomes thermally dissociated, also driving TiO to low abundances. A pressure-temperature profile of our generic hot Jupiter orbiting at 0.025 AU shows that TiO never reaches very high abundnaces below 10 mbar. This shows that TiO is not the cause of the inversion.  \label{fig:tio_pp}}
\end{figure}

\begin{figure}[ht]
	\center    \includegraphics[width=3.5in]{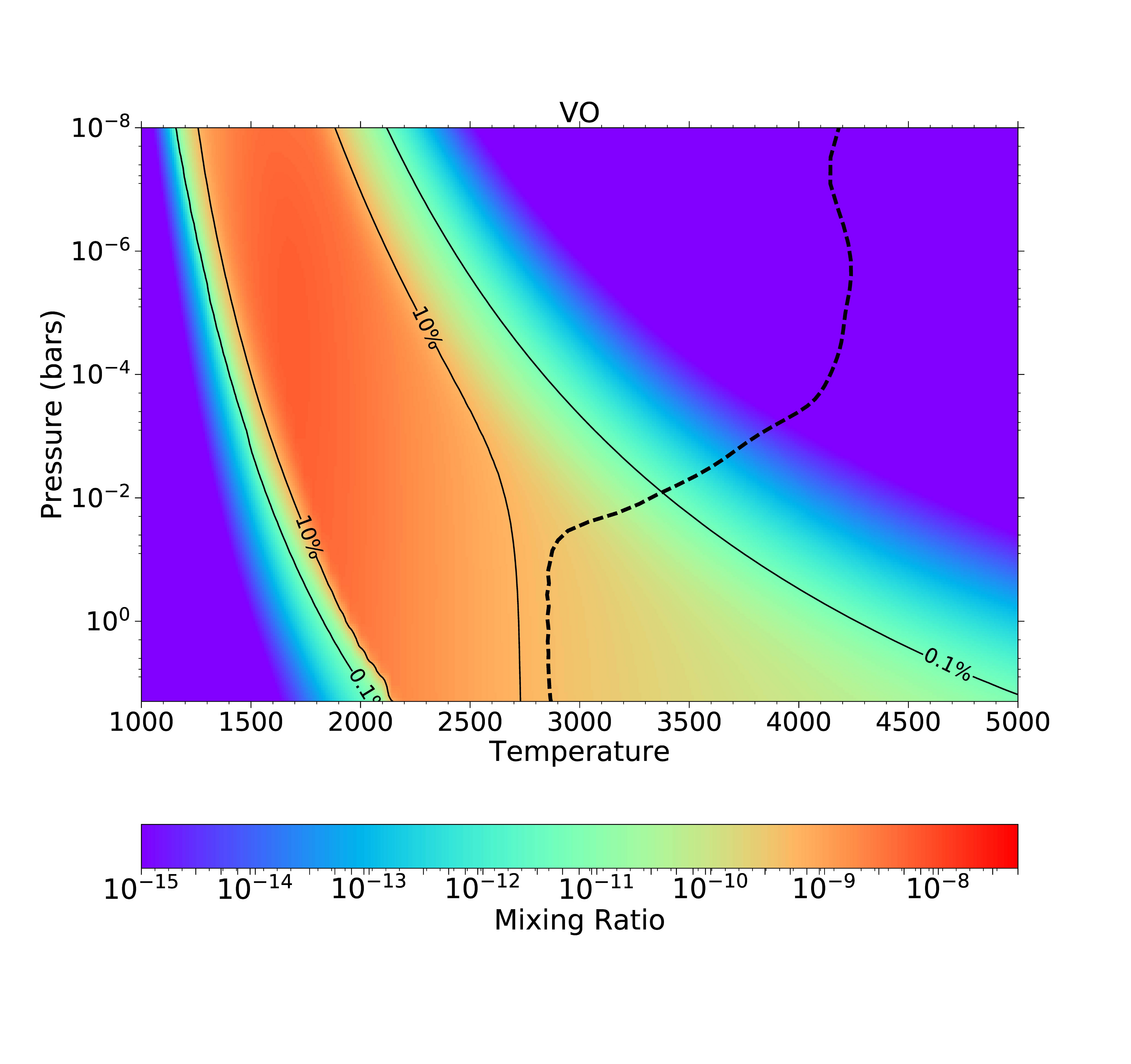}
	\vspace{-35pt}
	\caption{Same as Figure \ref{fig:tio_pp} but for VO. VO will be in high abundance at lower temperatures relative to TiO but also thermally dissociates at somewhat lower temperatures than TiO. In particular, even at pressures above 10 mbar, VO is depleted above 2700 K. \label{fig:vo_pp}}
\end{figure}

Figures \ref{fig:tio_pp} and \ref{fig:vo_pp} show the mixing ratio of TiO and VO as a function of pressure and temperature. Overplotted is the temperature profile of the planet at 0.025 AU. The planet reaches such high temperatures in its inversion that both TiO and VO are thermally dissociated. This implies that there are other opacity sources causing the thermal inversion seen at temperatures ${>}$2500~K and is the reason why TiO and VO become irrelevant for the highest temperature models in Figure \ref{fig:tp_sma}.

Similar trends of depletion can be seen when considering the atmosphere's atomic constituents. Hydrogen is only in its molecular form, H$_2$, above about 10 mbar. Below this pressure, hydrogen is in its atomic form. Since hydrogen is by far the most abundant element, this transition has a dramatic effect on properties like the scale height and specific heat (and therefore the adiabatic temperature gradient and radiative relaxation timescale). In planets with significant H$_2$ dissociation on the dayside, recombination of H back to H$_2$ at cooler longitudes can increase the efficiency of heat transport \citep{bell:2018}. Additionally, the transition from H$_2$ to H has a fundamental effect on atmospheric opacity due to the fact that the spectroscopically inactive diatomic molecule H$_2$ turns into a spectroscopically active form in atomic H. Similarly, the atmosphere will lose a significant continuous opacity in collision-induced absorption of H$_2$, although this will be compensated by the appearance of H$^-$ continuous opacity (see Section \ref{section:opac}). However, H$^-$ begins to become depleted below pressures of 1 mbar despite increasing abundances of both free electrons and H atoms. This is due to mutual neutralization with positive ions. Photodetachment may also remove a significant amount of H$^-$, but this is not included in our model.

Atomic species experience high rates of thermal ionization, with Na and K becoming ionized as deep as 100 mbar. Below 50 mbar, Na$^+$ and K$^+$ have replaced Na and K. Similarly, Fe, which we suggest is important in shaping the temperature structure through its absorption of short-wavelength irradiation, is mostly ionized around 0.5 mbar, at which points Fe$^+$ becomes the dominant form of Fe.

Figures \ref{fig:pp} and \ref{fig:pp_trace} make clear that the assumption of uniform vertical abundances in extremely irradiated hot Jupiters is incorrect. A non-detection of H$_2$O in the atmosphere of a planet may be the result of thermal dissociation rather than from non-solar elemental abundances. Thermal dissociation thus makes H$_2$O a poor measure of C/O ratio in extremely irradiated hot Jupiters. Importantly, other molecules are similarly dissociated, perhaps the most significant of which are TiO and VO. We discuss the consequences of this on the opacity and emission spectrum in Sections \ref{section:pdcf} and \ref{section:SE}, respectively.

\subsection{Opacity} \label{section:opac}

\begin{figure*}[ht]
	\center    \includegraphics[width=7.5in]{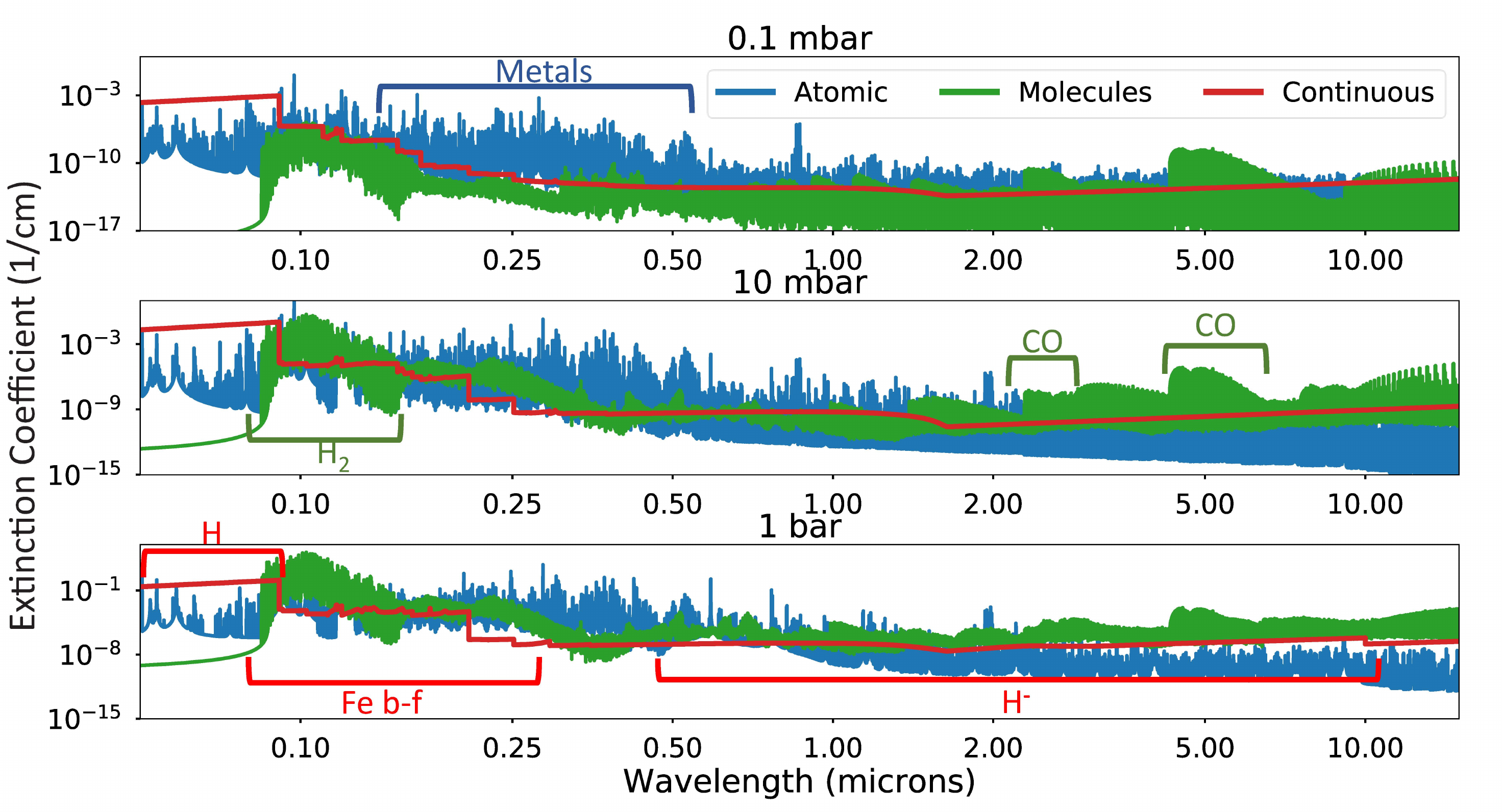}
	\caption{ The extinction coefficient (cm$^{-1}$) in the UV, optical, and near-IR as a function of wavelength from atomic, molecular, and continuous sources for the generic hot Jupiter model at 0.025 AU. Between 0.2 and 0.5 microns, where most of the stellar energy is located, atomic opacity, mostly from metal atoms like Fe, dominates. Shortward of 0.25 microns, bound-free opacity from Fe and H are important. Molecular absorption from electronic transitions of H$_2$, CO, and SiO can also be important at pressures 10 mbar and above.  Longward of 0.5 microns, continuous opacity, mainly from H$^-$, dominates at pressures of 10 mbar and below. \label{fig:opac}}
\end{figure*}

\begin{figure*}[ht]
	\center    \includegraphics[width=7.5in]{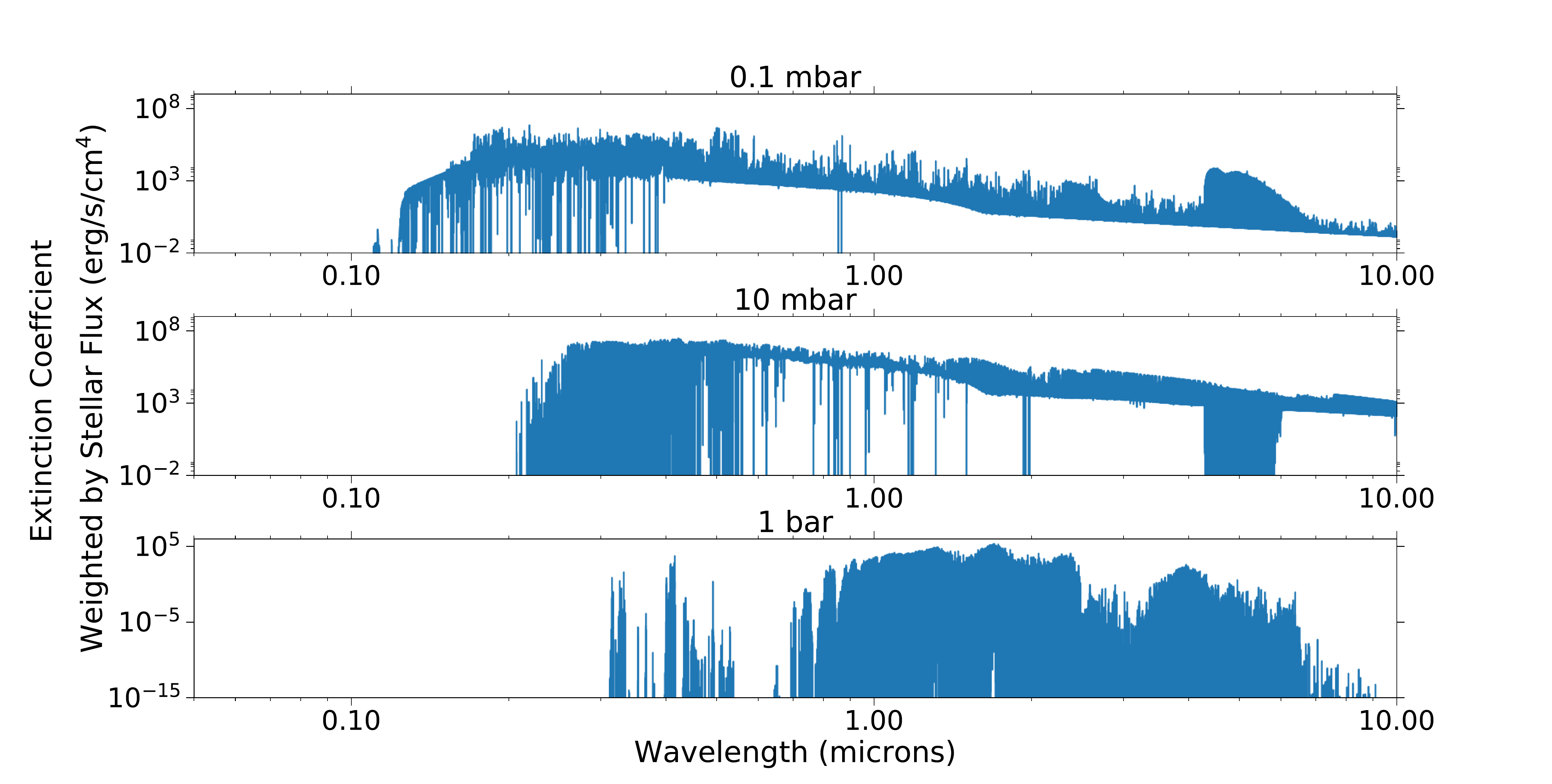}
	\caption{ The total atmospheric extinction coefficient weighted by the stellar flux at 0.1 mbar, 10 mbar, and 1 bar. Wavelengths where values are high indicated wavelengths where stellar flux is being absorbed. Wavelengths with low values indicate that either there is little opacity at that wavelength or there is little stellar flux at that wavelength and level. For example, at 1 bar, very little opacity is absorbing stellar flux shortward of 1 micron because most of the stellar flux has been absorbed at lower pressures. \label{fig:weightedopac}}
\end{figure*}

\begin{figure}[ht]
	\center    \includegraphics[width=3.75in]{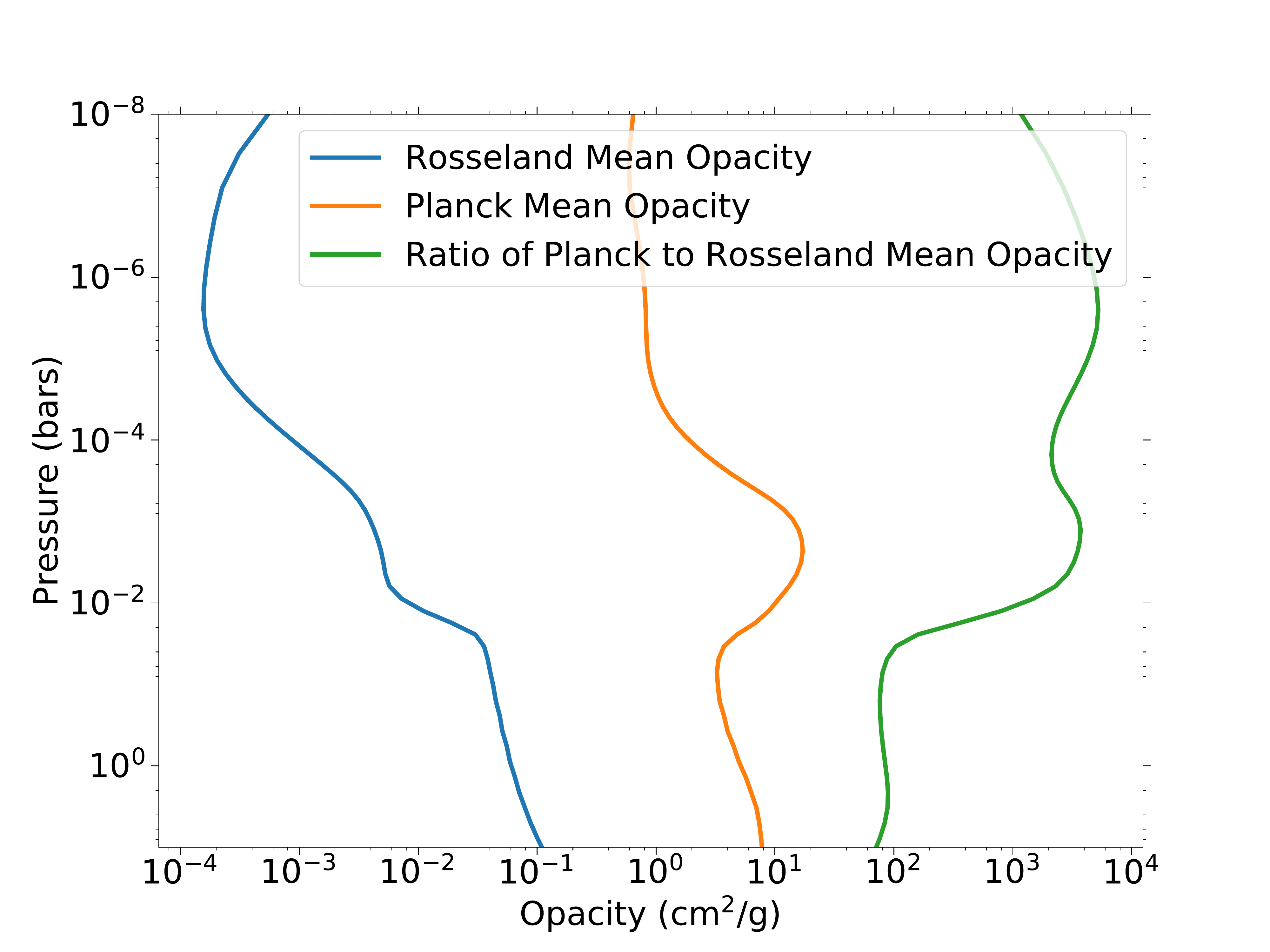}
	\caption{Rosseland (blue) and Planck (gold) mean opacities as a function of pressure for the generic hot Jupiter model at 0.025 AU. The ratio of the two mean opacities (green) is a measure of the non-grayness of the atmosphere.
  \label{fig:rmpm}}
\end{figure}

Having discussed the temperature structure and the molecular abundances, we now turn to the opacity structure in detail. Figure \ref{fig:opac} shows the extinction coefficient (cm$^{-1}$) for atomic, molecular, and continuous opacity at three different pressures, 0.1 mbar, 10 mbar, and 1 bar. As mentioned above, the temperatures at all parts of the dayside temperature structure for the model at 0.025 AU are above the condensation of clouds so we do not include condensate opacity. 

At 0.1 mbar, in the middle of the inversion where the maximum temperatures are reached, atoms are the main absorber at wavelengths shorter than 0.5 microns, which is where the majority of the incoming stellar flux is present. Molecular opacity is so low that continuous opacity dominates in most of the rest of the spectrum. The electronic transitions of H$_2$ are the only short-wavelength molecular opacities important at low pressures. The only other significant source of molecular opacity comes from CO, absorbing at its fundamental roto-vibrational band at 4.67 microns and its first overtone band at 2.3 microns. 

The large continuous opacities shortward of 912 \AA \xspace are from the bound-free transitions of H. Continuous opacity between 912 and 2500 \AA \xspace come mainly from bound-free Fe transitions. At longer wavelengths, the continuous opacity is dominated by H$^-$ opacity. Recently, \cite{arcangeli:2018} highlighted the importance of H$^-$ opacity in extremely hot exoplanet atmospheres, though its importance has been known in the brown dwarf and stellar community for quite some time \citep{wildt:1939,chandra:1945,chandra:1960:radtran,lenzuni:1991,sharp:2007,freedman:2008,freedman:2014}. We discuss the consequences of H$^-$ opacity in Section \ref{section:hminus}.

Another way to visualize the opacity is shown in Figure \ref{fig:weightedopac}. This shows the extinction coefficient weighted by the stellar flux at 0.1 mbar, 10 mbar, and 1 bar, emphasizing only those opacities which are important for the absorption of the irradiation at a given level. This figure shows that opacity shortward of 0.1 microns is unimportant at pressures 0.1 mbar and higher because the incoming stellar irradiation at those wavelengths is small and has been absorbed higher up. As the pressures get larger, more short-wavelength flux has been absorbed by the layers above. At 1 bar, nearly all flux shorter than 1 micron has already been absorbed.

Figure \ref{fig:rmpm} shows the Planck mean opacity and the Rosseland mean opacity as a function of pressure. The Planck mean opacity is defined as

\begin{equation}
	\kappa_P= \frac{\int_{0}^{\infty} \kappa_{\lambda} B_{\lambda} d\lambda}{\int_{0}^{\infty} B_{\lambda} d\lambda}
\end{equation}

and the Rosseland mean opacity is defined as 

\begin{equation}
	\frac{1}{\kappa_R}= \frac{\int_{0}^{\infty}  \frac{1}{\kappa_{\lambda}} \frac{dB_{\lambda}}{dT} d\lambda}{\int_{0}^{\infty}  \frac{dB_{\lambda}}{dT} d\lambda}.
\end{equation}

While the Planck mean opacity is an arithmetic mean weighted by the local Planck function, the Rosseland mean opacity is a harmonic mean weighted on the derivative of the local Planck function. The major contributor to the Planck mean opacity are opacity maxima near the peak of the local Planck function, while the major contributor to the Rosseland mean opacity are opacity minima \citep{freedman:2014}. The ratio of the Planck to the Rosseland mean opacity quantifies how non-grey the atmosphere is behaving (i.e., how much non-grey effects are determining the temperature structure) \citep{king:1956,parmentier:2014}. When $\kappa_P/\kappa_R {\sim} 1$, this implies that opacity maxima and minima are comparable and thus the opacity structure of the atmosphere does not exhibit much dynamic range. Figure \ref{fig:rmpm} shows that $\kappa_P/\kappa_R \gg 1$ for all parts of the atmosphere, therefore non-grey effects dominate in extremely irradiated atmospheres and grey approximations will result in poor estimates of atmospheric properties.

\subsubsection{Stellar Flux Penetration Depth and Contribution Function} \label{section:pdcf}

\begin{figure*}[ht]
	\center    
	\includegraphics[width=7.5in]{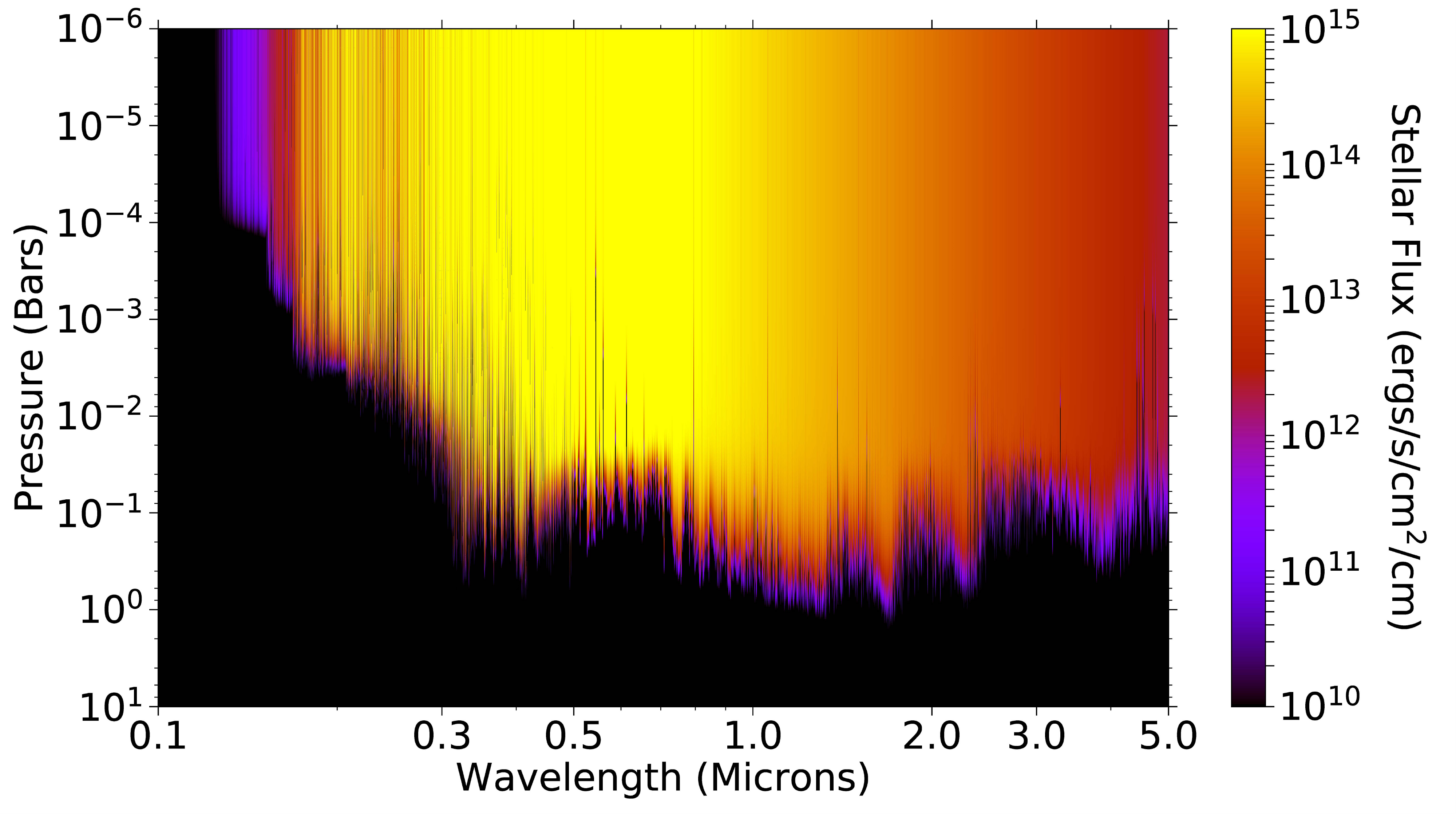}
	\caption{The stellar flux penetration as a function of wavelength (Eq. 3) for the generic hot Jupiter model at 0.025 AU. The pressures at which the stellar flux transitions from red to black indicates areas of absorption. Irradiation between 0.2 and 0.5 microns is absorbed between 1-100 mbar, driving heating in these layers. \label{fig:pd}}
\end{figure*}

\begin{figure*}[ht]
	\center    
	\includegraphics[width=7.5in]{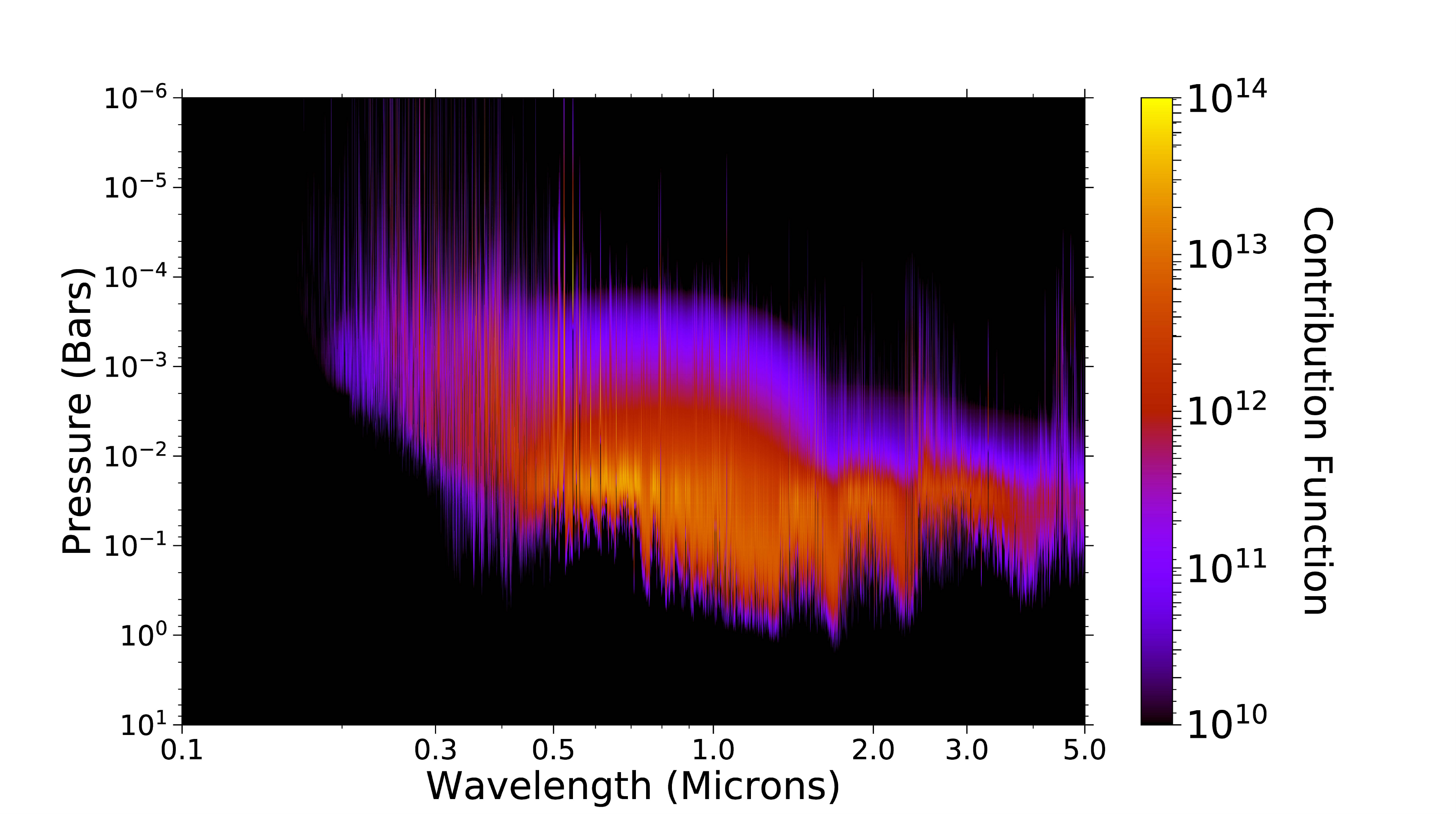}
	\caption{The contribution function as a function of wavelength (Eq. 4) for the generic hot Jupiter model at 0.025 AU. At locations where H$_2$O absorbs, namely 1.4 and 1.85 microns, the contribution function indicates outgoing flux comes from an isothermal level and is not differentiated much from the continuum. Additionally, there is not enough H$_2$O at lower pressures to bring the contribution function into the inversion layer because of thermal dissociation. This can be compared with the CO band heads at 2.3 and 4.65 microns that show outgoing flux coming from lower pressures, and thus higher temperatures, than the continuum. The effect of H$^-$ bound-free continuous opacity shortward of 1.6 microns mutes spectral features in that region. \label{fig:cf}}
\end{figure*}

To better understand the wavelengths at which absorption is important due to the incoming stellar irradiation, Figure \ref{fig:pd} shows the stellar flux penetration depth as a function of wavelength. We define the stellar flux penetration depth as:

\begin{equation}
PD(p,\lambda) = F_{p,\lambda}*e^{-\tau _{p,\lambda}}.
\end{equation}

\noindent This quantity describes the stellar flux passing through any given pressure level. Pressures where this quantity rapidly decrease are regions where the stellar flux is absorbed. Figure \ref{fig:pd} shows that the majority of the incoming stellar irradiation is being absorbed between 10 and 100 mbar. Much of the stellar flux shortward of 0.5 microns is being absorbed higher in the atmosphere, driving the inversion. Some strong lines in the optical absorb stellar flux at significantly lower pressures. In cases where TiO and VO absorption is important, flux between 0.5 and 1 micron would absorb higher in the atmosphere.

To better understand from what pressures flux is being emitted from, Figure \ref{fig:cf} shows the contribution function of the atmosphere, defined as:

\begin{equation}
CF(p,\lambda) = B_{\lambda}*e^{\tau _{p,\lambda}} \frac{d\tau _{p,\lambda}}{dp}.
\end{equation}

\noindent Between 0.5 and 1.6 microns, the lowest pressure that contribute to the outgoing flux is determined by the H$^-$ opacity, essentially raising the photosphere of the planet. Beyond 1.6 microns, the lower pressure limit of the contribution function is relatively isobaric except for at the CO bandheads. This is caused by the thermal inversion, whose high temperatures destroy the molecules that would otherwise be sculpting the contribution function at infrared wavelengths. 

The highest pressures that contribute to the outgoing flux are determined by molecular opacity from H$_2$O, CO, and CO$_2$. This implies that molecular opacity affects the opacity structure at higher pressures, since this is where the molecules are still high in abundance. However, as we discuss in Section \ref{section:SE}, the areas where H$_2$O and CO$_2$ opacity exist are isothermal and at lower pressures these molecules are thermally dissociated, resulting in the planet's emission spectrum being devoid of H$_2$O and CO$_2$ features.

\section{Discussion} \label{section:discussion}

\subsection{Ion Production} \label{section:ions}
Ionization plays an important role in hot Jupiter atmospheres. First, the ionization of alkali atoms like Na and K can result in a detectable decrease in the alkali abundance relative to neutral chemistry expectations. Perhaps the clearest example is HD 209458b. \cite{charbonneau:2002} used medium resolution HST/STIS observations to measure Na absorption in transit. The measured depth was about 3 $\times$ shallower than theoretical expectations. In addition to cloud opacity and non-LTE effects, photoionization has been suggested as an explanation for the lower-than-expected Na abundance. As described in \cite{barman:2002} and \cite{fortney:2003}, the magnitude of the effect of ionization on observations like transit spectroscopy, depends on the photoionization depth (i.e., the depth in the atmosphere where ionization by stellar photons stops playing a significant role in atmosphere chemistry).

As mentioned in Section \ref{section:abundances}, in our generic hot Jupiter model, thermal ionization of Na is significant as deep as 100 mbar and by 50 mbar sodium is mostly ionized. A similar effect is seen for K. Photoionization can drive ionization even deeper. This explains the relative unimportance of alkali atomic opacity on the absorption of stellar irradiation and on the resulting temperature structure (see Figure \ref{fig:fe_tp}). We predict that alkali absorption will be very muted or entirely absent in extremely irradiated hot Jupiters due to ionization. Cooler temperatures at a planet's terminators may allow some degree of recombination, but this will depend on the efficiency of temperature redistribution. 

The other important effect of ionization is the creation of ions and electrons which can experience Lorentz forces in the presence of the planet's magnetic field. This additional force will have an appreciable effect on the atmospheric dynamics and circulation \citep{koskinen:2014,rogersandshowman:2014,rogers:2017}. In some hydrodynamic models (non-magnetohydrodynamic), these Lorentz forces are added as a frictional drag force. \cite{tad:2016} and \cite{tad:2017} showed that this drag force plays a role in the measured day-night temperature differences, though perhaps secondary to the radiative timescale. Strong drag will effectively increase the advective timescale (i.e., the timescale at which a parcel of air can advect its heat away) and when the advective timescale is larger than the radiative timescale, large day-night temperature differences result. When drag occurs on timescales less than the rotation rate, the importance of drag becomes more important, efficiently quelling zonal winds \citep{tad:2016}. It is thus predicted that atmospheres with higher temperatures will experience more drag and a shorter radiative time scale, increasing the day-night temperature contrast. Observation of this trend is still tentative and may not be present at current precision \citep{tad:2017,parmentier:2017}.

Oscillatory behavior in the planetary winds can occur due to the coupling of the drag force and the planetary magnetic field \citep{rogersandtad:2014}. This may be responsible for the shift in observed phase curve offsets in HAT-P-7b, which was observed with \textit{Kepler} during the entirety of its prime mission \citep{Armstrong:2016,rogers:2017}. Additionally, when the ion fraction is high, either through photoionization or thermal ionization, atmospheric dynamos can form, significantly altering the behavior of the planetary magnetic field \citep{rogers:2017b}. \cite{batygin:2013} found that dipole magnetic fields can lead to latitudinally and radially non-uniform forces. If the magnetic field is in any way asymmetric or misaligned with the rotation axis, it may be possible to create latitudinally and longitudinally asymmetric dynamical flows.

These same Lorentz forces are what drives Ohmic dissipation in hot Jupiters, which may be responsible for the inflation of radii seen in many hot Jupiters \citep{batygin:2010}. Inflated radii also seem to be more common among the most highly irradiated planets \citep{guillot:2002,laughlin:2011,demory:2011c,miller:2011}. Recently, \cite{thorngren:2017} showed that the distribution of hot Jupiter masses and radii is consistent with the inflation efficiency predicted by Ohmic dissipation. This inflation efficiency increases with incident flux to increasing ionization but then starts to decrease after reaching a maximum at T$_{eq} \approx 1500$~K as magnetic drag forces begin to slow atmospheric wind speeds.

Due to the strong thermal inversion we find in the middle atmosphere of our generic extremely irradiated hot Jupiter, $e^-$ mixing ratios are as high as $10^{-4}$, which is about 3 orders of magnitude higher than the 1900 K model from \cite{rogersandtad:2014}. Thus the strong inversions in our models serve to increase thermal ionization at low pressures and may have significant effects on the circulation, magnetic field, and internal structure of extremely irradiated hot Jupiters. For the models presented here, the $e^-$ mixing ratio only reflects thermal ionization, so adding photoionization into the chemical equilibrium solution would increase the $e^-$ mixing ratio further.

\subsection{Non-Local Effects}

For the models we have presented above, we have assumed local thermodynamic equilibrium (LTE). One part of this assumption is that when radiation gets absorbed by a particle in the atmosphere, that particle has time to thermalize that energy with the rest of the atmosphere through collisions. When collisions dominate, we can assume a Maxwellian distribution of states. This means that by knowing the temperature, we know the distribution of level populations for the atoms and molecules. However, at low pressures and in the presence of strong irradiation, radiative rates can become greater than corresponding collisional rates. When this happens, atoms and molecules are no longer in LTE, the distribution of states is no longer Maxwellian, and calculating how the atoms and molecules radiate becomes much more complex.

Non-LTE effects will change the opacity and temperature structure of an atmosphere. While PHOENIX can calculate both the departure coefficients and temperature structure simultaneously and self-consistently, convergence is computationally intensive and difficult. This is in part due to the complex and sensitive numerical calculations that must be done. Additionally, the main difficulty is a lack of collisional data for many atmospheric constituents. For example, the collisional rates between atomic Na and $e^-$, H, and He are known \citep{belyaev:2010,lin:2008}, but collisional rates between Na and H$_2$ are not known due to the asymmetry of the H$_2$ nucleus.

While full non-LTE calculations are beyond the scope of this paper, we do note that in our attempts to self-consistently model the departure coefficients and temperature structure, the inversion found in our models is amplified and does not go away. 

Photochemistry is another non-local effect that is thought to be important in exoplanet atmospheres. Photochemistry is not included in this version of PHOENIX; however, the extreme temperatures particularly in the upper atmosphere of these planets would prevent photochemical products from building up. Additionally, only extremely vigorous vertical mixing would allow important photochemical sources like CH$_4$ to exist in appreciable amounts in the middle and upper atmosphere of these planets.

\subsection{Secondary Eclipse Spectra} \label{section:SE}

\begin{figure*}[ht]
	\center    
	\includegraphics[width=7in]{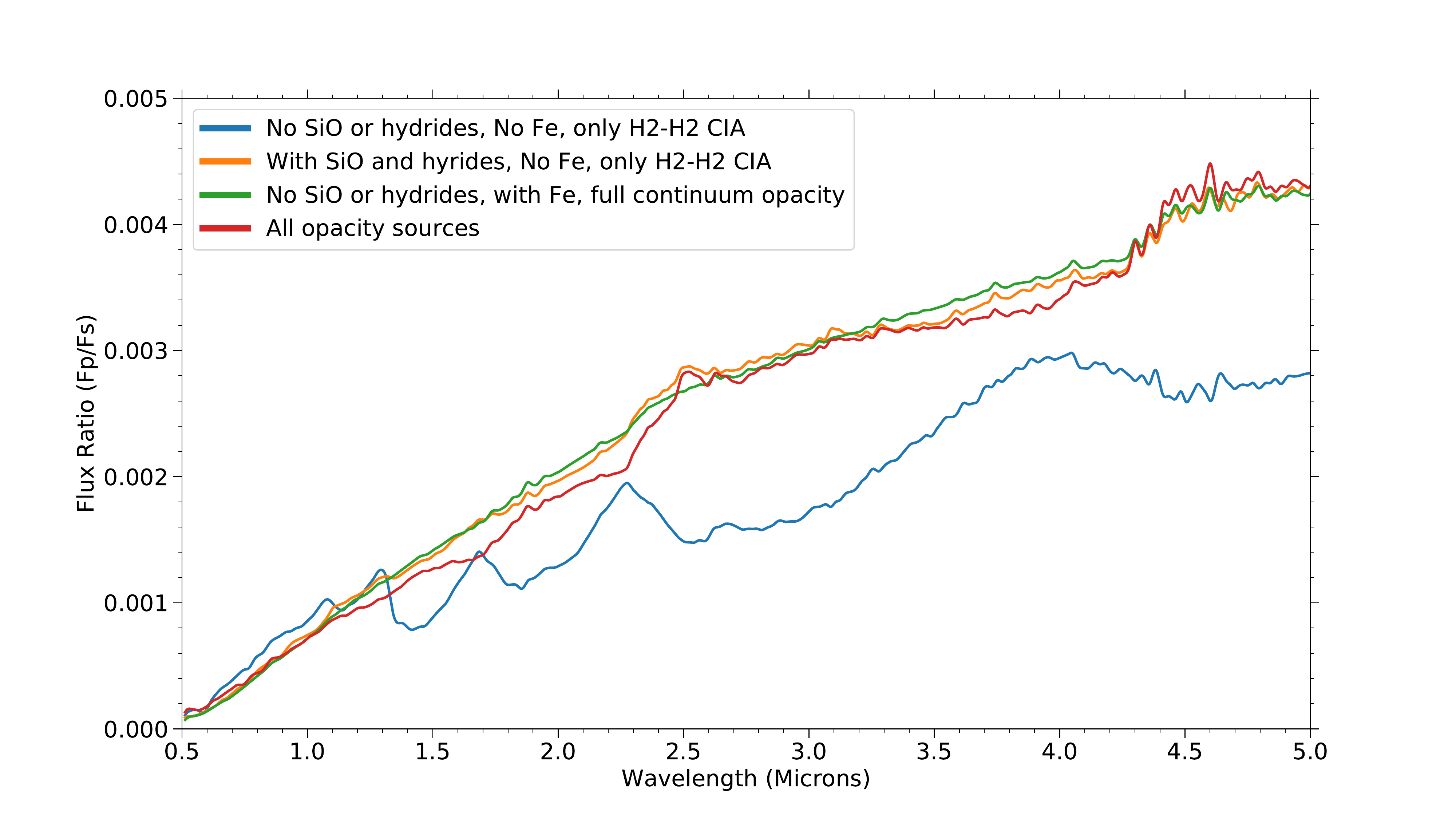}
	\caption{Secondary eclipse spectra of the models described in Figure \ref{fig:fe_tp}. Model 1 (blue) is the only model to show large absorption features since it is the only model without a temperature inversion. Since the non-inverted model exhibits cooler temperatures at low pressures, the molecules remain abundant and are not thermally dissociated. All other models show no large H$_2$O features but do show CO in emission at 2.3 and 4.65 microns. The large temperatures in the inversions in these models thermally dissociate water at lower pressures and create large amounts of H$^-$ opacity that obscures the water feature at lower altitudes (see Section \ref{section:hminus}). \label{fig:spectra}}
\end{figure*}

We predict that the dayside spectrum of extremely hot Jupiters will look qualitatively and quantitatively different from lower temperature hot Jupiters. Figure \ref{fig:spectra} shows simulated secondary eclipse spectra for the same temperatures structures in Figure \ref{fig:fe_tp}. The inverted spectra (gold, red, and green) are clearly much different from the non-inverted model (blue). While the non-inverted model is shaped by H$_2$O and CO absorption, the inverted models do not show evidence of H$_2$O and show CO in emission. As mentioned in Sections \ref{section:abundances} and \ref{section:pdcf}, H$_2$O opacity is only significant where the atmosphere is isothermal. At lower pressures, H$_2$O thermally dissociates due to the thermal inversion. 

Again, we stress that this scenario is true for more than just H$_2$O, in fact most molecules will exhibit this behavior, with the exception of CO, which can exists at lower pressures and higher temperatures than all other molecules (see Section \ref{section:abundances}). We predict that if H$_2$O spectral features are absent, then TiO and VO features will be similarly missing.

\subsubsection{The Role of H$^-$}\label{section:hminus}

\begin{figure*}[ht]
	\center    \includegraphics[width=7in]{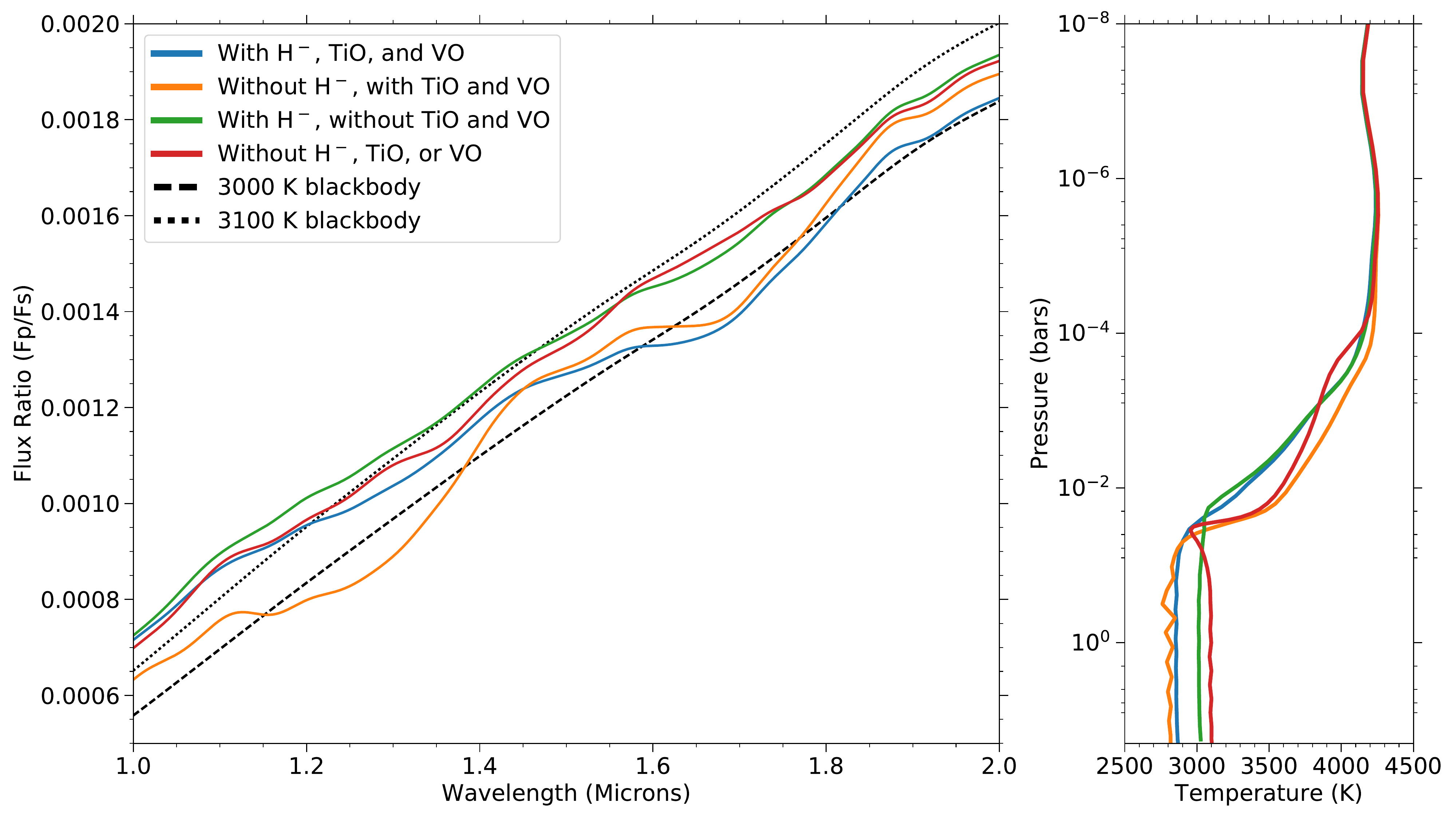}
	\caption{Left: The secondary eclipse spectrum of our generic hot Jupiter for different scenarios involving H$^-$ opacity. The blue model is our fiducial model with all opacity sources included. Right: The temperature pressure profiles for the same models. \label{fig:hminus}}
\end{figure*}

Previous work has pointed out the significance of H$^-$ opacity at high temperatures \citep{wildt:1939,chandra:1945,chandra:1960:radtran,lenzuni:1991,sharp:2007,freedman:2008,freedman:2014}. \cite{arcangeli:2018} recently argued that the combined effects of H$^-$ opacity and thermal dissocation are responsible for the lack of H$_2$O features in the secondary eclipse spectrum of WASP-18b. Figure \ref{fig:hminus} shows model scenarios with and without H$^-$ opacity for our generic hot Jupiter. H$^-$ continuous opacity increases the optical depth at all wavelengths, raising the photosphere to lower pressures. This results in the temperature inversion occurring at lower pressures as well.

For the case where TiO and VO are present (blue and gold models), it is clear that H$^-$ opacity is serving to mute the H$_2$O feature at 1.4 microns. Interestingly, H$^-$ mutes the short-wavelength half of the H$_2$O feature more strongly than the long-wavelength half. This is due to the fact that H$^-$ is reaching an opacity minimum near 1.6 microns. 

However, in our models without TiO and VO (green and red models), H$_2$O features are absent in the secondary eclipse spectrum regardless of whether H$^-$ opacity is included in the model or not. Models without TiO and VO are 100-200 K warmer at high pressures than models with TiO and VO due to the fact that the stellar irradiation can heat higher pressures in the atmosphere in the absence of TiO and VO. In our generic hot Jupiter model, these higher temperatures are enough for H$_2$O to be thermaly dissociated throughout more of the atmosphere. Thus our models without TiO and VO will not show H$_2$O features mostly because of thermal dissociation, not H$^-$ opacity.

\begin{figure*}[ht]
	\center    \includegraphics[width=7in]{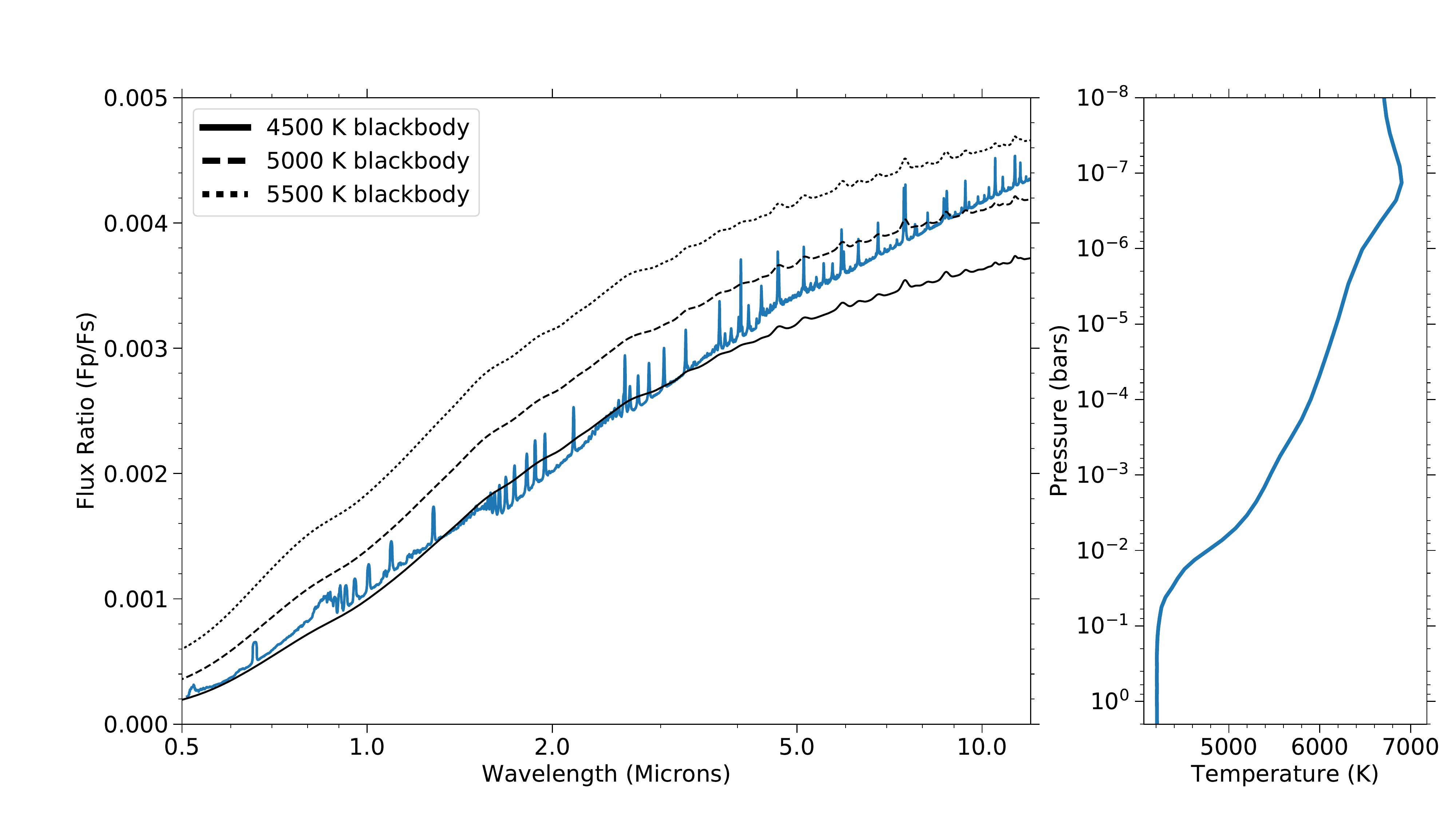}
	\caption{Left: The secondary eclipse spectrum of KELT-9b, assuming dayside redistribution. Nearly all molecules have been thermally dissociated by 100 mbar. The main opacity source is from H$^-$. Most of the lines seen in the eclipse spectrum are from absorption lines in the stellar spectrum. Right: The temperature pressure profile for the same model. \label{fig:k9obs}}
\end{figure*}

While H$^-$ opacity's effect on atmospheres like our generic hot Jupiter is important, at even higher temperatures H$^-$ opacity becomes the most significant opacity source across almost all infrared wavelengths. Figure \ref{fig:k9obs} shows a PHOENIX model of the secondary eclipse spectrum of KELT-9b, the hottest known Jovian exoplanet \citep{gaudi:2017}. While the spectrum is nearly devoid of molecular absorption or emission, the spectrum is not isothermal. The brightness temperature varies smoothly with wavelength due to H$^-$ opacity, whose free-free opacity increases with wavelength above 1.6 microns. Between 2 and 12 microns, the brightness temperature of KELT-9b increases by nearly 1,000 K. H$^-$ becomes important when hydrogen is in its atomic form and a supply of free electrons exist. For a planet like KELT-9b, atomic hydrogen is the most abundant species throughout the atmosphere until about a microbar when H$^+$ and $e^-$ become the main constituents.

\begin{figure*}[ht]
	\center    \includegraphics[width=6.5in]{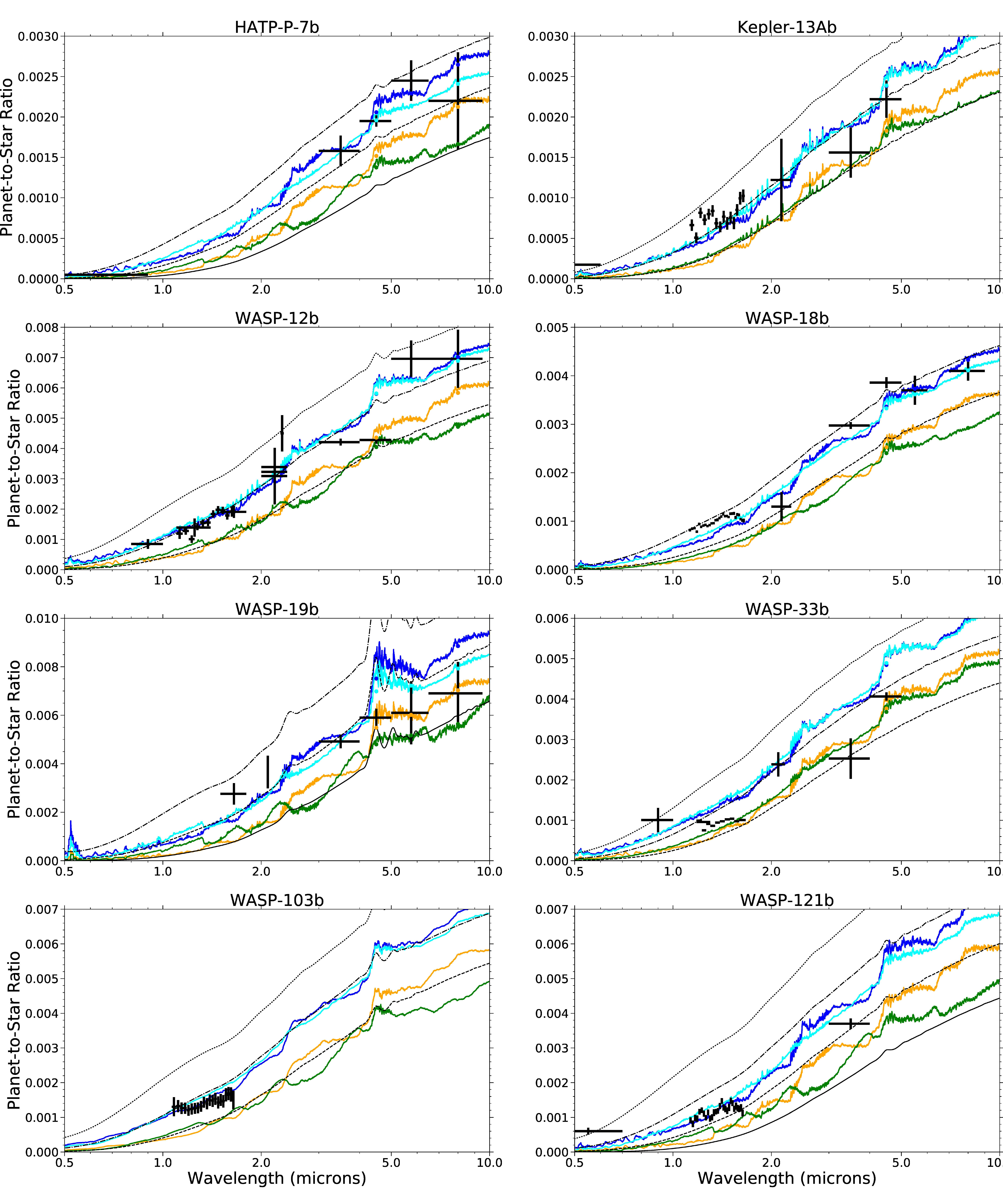}
	\caption{Previously published observations of extremely irradiated hot Jupiters. Blue models assume dayside temperature redistribution. Cyan models are the same as blue but neglect the presence of TiO and VO. Gold models assume full temperature redistribution. Green models are the same as gold but neglect the presence of TiO and VO. Black body spectra are plotted for temperatures of 2000, 2500, 3000, and 3500 K indicated by the solid line, dashed line, dash-dot line, and dotted line respectively. No attempt was made at fitting the models to the data. References are shown in Table \ref{table:properties}. \label{fig:previousobs}}
\end{figure*}

\begin{figure*}[ht!]
	\center    \includegraphics[width=7.5in]{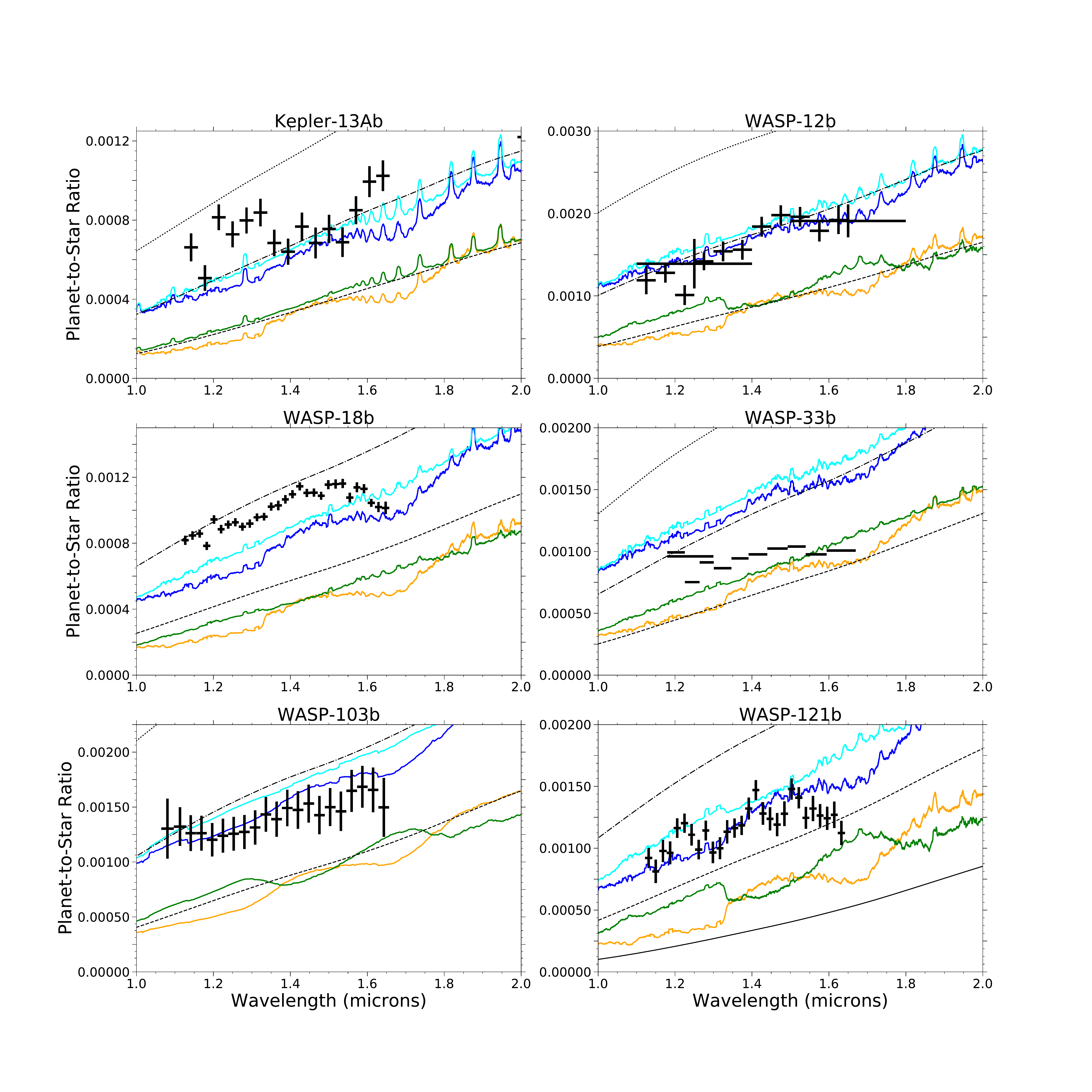}
	\caption{Same as Figure \ref{fig:previousobs}, but for planets with previous observations with HST/WFC3 G141. The legend is the same as Fig \ref{fig:previousobs}. Again, no attempt was made to fit the data. \label{fig:previouswfc3obs}}
\end{figure*}

\subsubsection{Comparison to Secondary Eclipse Observations} \label{section:pastobs}

Figure \ref{fig:previousobs} shows previous secondary eclipse observations of eight hot Jupiters with equilibrium temperatures greater than 2,000 K. The planetary parameters used and references for the observations are listed in Table \ref{table:properties}. Most of the data comes from HST/WFC3 and Spitzer. Figure \ref{fig:previousobs} also shows four different model scenarios: full temperature redistribution with TiO and VO, full temperature redistribution without TiO and VO, dayside only temperature redistribution with TiO and VO, and dayside only redistribution without TiO and VO. No attempt was made to fit the data, beyond choosing the planetary parameters.

While the error bars for the Spitzer points in Figure \ref{fig:previousobs} are relatively large, they are in general agreement with at least one of our model scenarios. The biggest exception to this are the 3.6 and 4.5 micron Spitzer points for WASP-12b from \cite{stevenson:2014}. These points are both much lower than one would expect from our models. The low 4.5 micron Spitzer points could indicate absorption features of carbon species, but this is not expected in any model scenarios we have investigated.

For planets that have been observed with HST/WFC3, Figure \ref{fig:previouswfc3obs} shows the spectra and data in the G141 region (1.1-1.7 microns) in more detail. Both \cite{haynes:2015} and \cite{evans:2017} have interpreted the HST/WFC3 eclipse spectra of WASP-33b and WASP-121b, respectively, to show emission of H$_2$O. Strong VO emission was also claimed in WASP-121b, however the retrieval indicated abundances of about 1000 times solar metallicity.  In contrast, WASP-12b and WASP-18b, and WASP-103b have been interpreted as being devoid of any water spectral features \citep{stevenson:2014,sheppard:2017,arcangeli:2018,cartier:2017}. This lack of H$_2$O features has been initially interpreted as evidence for high C/O \citep{stevenson:2014,sheppard:2017}, however, as has been shown in this paper and \cite{arcangeli:2018}, thermal dissociation, H$^-$ opacity, and an isothermal deep atmosphere can also mask H$_2$O spectral features.

Kepler-13Ab is the only planet with claimed water absorption at 1.4 microns in this sample \citep{beatty:2017a}. This is not fit by our models, since we predict a strong inversion to form and H$_2$O to be thermally dissociate on the hot dayside that has a measured averaged dayside brightness temperature of 3000~K.

The longest wavelength points of the HST/WFC3 dayside spectrum of each of these planets, except for Kepler-13Ab, have a dip toward smaller planet-to-star flux ratio. This could in part be due to the fact that H$^-$, the dominant continuous opacity in hot planets, reaches its minimum opacity at 1.6 microns as mentioned in the previous section. This is consistent with the behavior of our models that include all opacity sources (i.e., the blue model in Figure \ref{fig:hminus}). Alternatively, this could be a common instrumental systematic behavior toward the edge of the detector.

 \begin{table*}[t] 
 	\centering  
 	\caption{Planet Properties For Comparison with Observations}
 	\label{table:properties} 
 	\begin{tabular}{p{1.8cm}p{2.0cm}p{1.1cm}p{1.1cm}p{1.3cm}p{1.5cm}p{4.0cm}}
 		\hline
 		Planet & Equilibrium Temperature (K)\tablenotemark{1} & Radius ($R_J$) & Mass ($M_J$)& log(g) (cm s$^{-2}$) & Host Star Temperature & References\\
\hline   
  		HAT-P-7b & 2270-2700 & 1.491& 1.682 & 3.27 & 6440 & \cite{christiansen:2010,wong:2015} \\ 
 \hline    
  		Kepler-13Ab & 2550-3050 & 1.521 & 9.28 & 4.0 & 7650 & \cite{shporer:2014,esteves:2015,beatty:2017a} \\ 
\hline
 		WASP-12b  & 2580-3070 & 1.9 & 1.47 & 3.00 & 6360 &  \cite{lopez-morales:2010,croll:2011a,crossfield:2012d,foehring:2013,stevenson:2014}\\
\hline  
 		WASP-18b & 2400-2850 & 1.3  & 10.2 & 4.28 & 6400 & \cite{sheppard:2017,arcangeli:2018} \\
\hline 
 		WASP-19b  & 2100-2500 & 1.392 & 1.069 & 3.14 & 5568 &  \cite{anderson:2013,wong:2015} \\
\hline  
 		WASP-33b & 2700-3200 & 1.6 & 2.1 & 3.3 &
 		7430 & \cite{haynes:2015} \\
\hline  
 		WASP-103b & 2500-3000 & 1.646 & 1.47 & 3.2 & 6110 & \cite{cartier:2017} \\
\hline 
 		WASP-121b & 2350-2800  & 1.865& 1.183 & 2.93 & 6460 & \cite{evans:2017} \\ 
 	\end{tabular}
 	\tablenotetext{1}{The range in equilibrium temperature between planet-wide heat redistribution and dayside-only heat redistribution.}
 \end{table*}

\subsubsection{Future Observations with JWST}

JWST will be an ideal facility for the characterization of extremely irradiated hot Jupiters. While HST/WFC3 can only observe the water feature at 1.4 microns and \textit{Spitzer} only has two approximately micron-wide passbands at 3.6 microns and 4.5 microns, JWST will be capable of spectroscopy from 0.6 to 28 microns, with most exoplanet spectroscopy focusing on the wavelength range from 0.6 to 12 microns. In this range, many molecules have several roto-vibrational bandheads, including H$_2$O, CO$_2$, CO, CH$_4$, TiO, and VO. JWST will be capable of placing constraints on both the molecular abundance and temperature structure of exoplanet atmospheres \citep{greene:2016}.

Extremely irradiated hot Jupiters provide some of the best targets for characterization with JWST. Using the figure of merit as defined in \cite{zellem:2017} to quantify target observability, nearly all of the highest ranked targets are ultra-hot Jupiters. As mentioned in Section \ref{section:intro}, this is due to the fact that extremely irradiated hot Jupiters have inflated radii and hot dayside atmospheres. Most are also likely too hot to possess clouds on their dayside.

JWST will be able to test the models and predictions we have presented here. We summarize our predictions here:

\begin{itemize}
	\item Most, if not all, planets above 2500 K will have temperature inversions.
	\item Thermal dissocation will mute most molecular spectral features, including H$_2$O, TiO, and VO, in planets above 2500 K, but CO will remain in emission at higher temperatures.
	\item The dayside spectrum of KELT-9b will be devoid of molecular features and will be dominated by continuous H$^-$ opacity.
\end{itemize}

\subsection{High-Dispersion Spectroscopy}

Since we predict that atomic lines become important in extremely hot Jupiters, a clear path forward towards characterization would include observations of these lines. However, resolving individual lines with current low- and medium-resolution exoplanet observing practices and instrumentation is difficult. JWST will reach maximum resolutions of $R\sim$3500 with NIRSpec, which is too low to observe small, thin atomic lines. Ground-based high-dispersion spectroscopy (HDS) allows exoplanets to be observed at high resolution by spectroscopically separating the planet's flux contribution from its host star and telluric lines in a way very similar to techniques used to detect spectroscopic binary stars \citep[e.g.,][]{snellen:2008,birkby:2013,dekok:2013,brogi:2014,hschwarz:2015}. A single individual line contains too low of a signal with current instrumentation so molecular band-heads, which consist of millions and sometimes even billions of lines are often targeted. With large wavelength coverage, ift may be possible to detect species like Fe or Mg in the atmosphere of extremely irradiated hot Jupiters since they can have many lines as well. While the planet-to-star flux ratio at optical wavelengths is very small, \cite{nugroho:2017} demonstrated that characterization at optical wavelengths is possible with HDS. Inaccuracies in the short-wavelength line list information of at least TiO will also need to be taken into account \citep{hoeijmakers:2015}. 

As discussed above in Section \ref{section:SE}, CO will be found in emission, even at low resolutions. However this signal will also be detectable with HDS at higher resolutions, providing independent verification and prospects for advanced characterization utilizing both techniques, similar to \cite{brogi:2017}.

\section{Conclusion} \label{section:conclude}

Extremely irradiated hot Jupiters provide some of the best observing targets for future characterization due to their large scale heights, short periods, and likely absence of clouds. However, their extreme temperatures stretch the capability of models designed for cooler objects. 

Using self-consistent PHOENIX models with opacity sources often not included in other models, we find the following:

\begin{itemize}
	\item Temperature inversions exist at pressures probed by secondary eclipse observations for planets ${>}$~2500 K regardless of the presence of TiO or VO due to a combination of short wavelength irradiation from early-type host stars and short wavelength absorption by continuous opacity, metal atoms, SiO, and metal hydrides.
	
	\item These high-temperature inversions lead to most molecules becoming thermally dissociated around 10-100 mbar, depriving the atmosphere of important sources of cooling. Retrieval analyses that assume uniform vertical abundances will consequently be biased towards sub-solar molecular abundances. 
	
	\item We predict future observations in secondary eclipse and with high-dispersion spectroscopy will show a lack of molecular features. One exception may be CO, the strongest molecule in nature, which will survive at higher temperatures than other molecules.
	
\end{itemize}

Using the predicted yield from the Transiting Exoplanet Survey Satellite (TESS) \citep{barclay:2018}, we find that TESS will discover about 81 planets with radii greater than Jupiter's, equilibrium temperatures above 2000 K (assuming planet-wide temperature redistribution), and K magnitudes greater than 13. This will increase the known population of characterizable extremely irradiated hot Jupiters by nearly a factor of 5. We suggest these objects as targets for future characterization, as they present many fundamental questions in planetary atmospheric physics while also being some of the most amenable targets to study. With facilities like HST, JWST, and the future generation of extremely large telescopes, we will be able to better understand the extraordinary atmospheres of this unique class of astrophysical object.

\acknowledgments

We thank Jayne Birkby and Ian Crossfield for useful discussions regarding extreme hot Jupiters. We also thank Sarah Peacock for providing the M0 dwarf temperature profile. This research was partially supported under programs HST-GO-12511 and HST-GO-14797 with financial support provided by NASA through a grant from the Space Telescope Science Institute, which is operated by the Association of Universities for Research in Astronomy, Inc., under NASA contract NAS 5-26555. This research has made use of the NASA Astrophysics Data System and the NASA Exoplanet Archive, which is operated by the California Institute of Technology, under contract with the National Aeronautics and Space Administration under the Exoplanet Exploration Program. Resources supporting this work were also provided by the NASA High-End Computing (HEC) Program through the NASA Advanced Supercomputing (NAS) Division at Ames Research Center. Allocation of computer time from the UA Research Computing High Performance Computing (HPC) at the University of Arizona is also gratefully acknowledged.

\bibliographystyle{apj}

\end{document}